\documentclass[
  man,
  floatsintext,
  longtable,
  nolmodern,
  notxfonts,
  notimes,
  colorlinks=true,linkcolor=blue,citecolor=blue,urlcolor=blue]{apa7}

\usepackage{amsmath}
\usepackage{amssymb}

\RequirePackage{longtable}
\RequirePackage{threeparttablex}

\makeatletter
\renewcommand{\paragraph}{\@startsection{paragraph}{4}{\parindent}%
	{0\baselineskip \@plus 0.2ex \@minus 0.2ex}%
	{-.5em}%
	{\normalfont\normalsize\bfseries\typesectitle}}

\renewcommand{\subparagraph}[1]{\@startsection{subparagraph}{5}{0.5em}%
	{0\baselineskip \@plus 0.2ex \@minus 0.2ex}%
	{-\z@\relax}%
	{\normalfont\normalsize\bfseries\itshape\hspace{\parindent}{#1}\textit{\addperi}}{\relax}}
\makeatother

\usepackage{longtable, booktabs, multirow, multicol, colortbl, hhline, caption, array, float, xpatch}
\usepackage{tcolorbox}
\tcbuselibrary{listings,theorems, breakable, skins}
\usepackage{fontawesome5}

\definecolor{quarto-callout-color}{HTML}{909090}
\definecolor{quarto-callout-note-color}{HTML}{0758E5}
\definecolor{quarto-callout-important-color}{HTML}{CC1914}
\definecolor{quarto-callout-warning-color}{HTML}{EB9113}
\definecolor{quarto-callout-tip-color}{HTML}{00A047}
\definecolor{quarto-callout-caution-color}{HTML}{FC5300}
\definecolor{quarto-callout-color-frame}{HTML}{ACACAC}
\definecolor{quarto-callout-note-color-frame}{HTML}{4582EC}
\definecolor{quarto-callout-important-color-frame}{HTML}{D9534F}
\definecolor{quarto-callout-warning-color-frame}{HTML}{F0AD4E}
\definecolor{quarto-callout-tip-color-frame}{HTML}{02B875}
\definecolor{quarto-callout-caution-color-frame}{HTML}{FD7E14}

\usepackage{hyperref}

\providecommand{\tightlist}{%
  \setlength{\itemsep}{0pt}\setlength{\parskip}{0pt}}
\usepackage{longtable,booktabs,array}
\usepackage{calc} 
\usepackage{etoolbox}
\makeatletter
\patchcmd\longtable{\par}{\if@noskipsec\mbox{}\fi\par}{}{}
\makeatother
\IfFileExists{footnotehyper.sty}{\usepackage{footnotehyper}}{\usepackage{footnote}}
\makesavenoteenv{longtable}

\usepackage{graphicx}
\makeatletter
\newsavebox\pandoc@box
\newcommand*\pandocbounded[1]{
  \sbox\pandoc@box{#1}%
  \Gscale@div\@tempa{\textheight}{\dimexpr\ht\pandoc@box+\dp\pandoc@box\relax}%
  \Gscale@div\@tempb{\linewidth}{\wd\pandoc@box}%
  \ifdim\@tempb\p@<\@tempa\p@\let\@tempa\@tempb\fi
  \ifdim\@tempa\p@<\p@\scalebox{\@tempa}{\usebox\pandoc@box}%
  \else\usebox{\pandoc@box}%
  \fi%
}
\def\fps@figure{htbp}
\makeatother

\NewDocumentCommand\citeproctext{}{}
\NewDocumentCommand\citeproc{mm}{%
  \begingroup\def\citeproctext{#2}\cite{#1}\endgroup}
\makeatletter
 \let\@cite@ofmt\@firstofone
 \def\@biblabel#1{}
 \def\@cite#1#2{{#1\if@tempswa , #2\fi}}
\makeatother
\newlength{\cslhangindent}
\newlength{\csllabelwidth}
\newenvironment{CSLReferences}[2] 
 {\begin{list}{}{%
  \setlength{\itemindent}{0pt}
  \setlength{\leftmargin}{0pt}
  \setlength{\parsep}{0pt}
  \ifodd #1
   \setlength{\leftmargin}{\cslhangindent}
   \setlength{\itemindent}{-1\cslhangindent}
  \fi
  \setlength{\itemsep}{#2\baselineskip}}}
 {\end{list}}
\usepackage{calc}

\usepackage{newtx}

\defaultfontfeatures{Scale=MatchLowercase}
\defaultfontfeatures[\rmfamily]{Ligatures=TeX,Scale=1}

\title{Extracting Bayesian Evidence from Frequentist \(p\)-Values}

\shorttitle{Extracting Evidence from \(p\)}

\usepackage{etoolbox}

\authorsnames[{1,3},{2},{3}]{Frederik Aust,Samuel Pawel,Eric-Jan
Wagenmakers}

\authorsaffiliations{
{Department of Psychology, University of Cologne},{University of
Zurich},{University of Amsterdam}}

\leftheader{Aust, Pawel and Wagenmakers}

\date{2026-07-13}

\abstract{The \(p\)-value and the Bayes factor are measures of evidence
that are often considered to be philosophically and mathematically
incompatible: The \(p\)-value quantifies conflict between data and
\(\mathcal{H}_0\) (``surprise''), whereas the Bayes factor quantifies
the relative predictive accuracy of \(\mathcal{H}_0\) versus
\(\mathcal{H}_1\) (``evidence''). We revisit Jeffreys's Approximate
Bayes factor (JAB)---a simple, largely overlooked approximation dating
back to the 1930s---which connects these two paradigms for objective
hypothesis testing of the existence of an effect. Under a
unit-information prior the approximation requires only the \(p\)-value
and the effective sample size \(n_\text{eff}\). We clarify the core
assumptions and boundary conditions for the application of JAB and show
across 704 published \(t\)-tests and 39 comparisons of proportions that
JAB approximates objective Bayes factors remarkably well. The connection
between \(p\)-values and JAB has a practical implication: The evidence
implied by a \(p\)-value depends strongly on \(n_\text{eff}\).
Conventional verbal labels for \(p\)-values (e.g., ``strong surprise''
for .001 \textless{} \(p\) \textless{} .01) correspond to similarly
graded Bayes factors only around \(n_\text{eff} \approx 8\); for larger
samples the same \(p\)-value implies weaker evidence. In moderately
sized to large samples, \(p > .10\) can amount to moderate or even
strong evidence for \(\mathcal{H}_0\). JAB offers a cheap,
sample-size-sensitive supplement to \(p\)-values, computable from
routinely reported statistics, that remains valid even under optional
stopping. }


\authornote{\par{\addORCIDlink{Frederik
Aust}{0000-0003-4900-788X}}\par{\addORCIDlink{Samuel
Pawel}{0000-0003-2779-320X}}\par{\addORCIDlink{Eric-Jan
Wagenmakers}{0000-0003-1596-1034}} 

\par{ Data and code to reproduce all results are available at
https://github.com/crsh/jabp.   No external funding was received for
this work.   }
\par{Correspondence concerning this article should be addressed
to Frederik Aust, Department of Psychology, University of
Cologne, Herbert-Lewin-Str.
2, Cologne 50931, Germany, Email: frederik.aust@uni-koeln.de}
}

\makeatletter
\let\endoldlt\endlongtable
\def\endlongtable{
\hline
\endoldlt
}
\makeatother

\usepackage{booktabs}
\usepackage{longtable}
\usepackage{array}
\usepackage{multirow}
\usepackage{wrapfig}
\usepackage{float}
\usepackage{colortbl}
\usepackage{pdflscape}
\usepackage{tabularx}
\usepackage{xltabular}
\usepackage{threeparttable}
\usepackage{threeparttablex}
\usepackage[normalem]{ulem}
\usepackage{makecell}
\usepackage{xcolor}
\newcommand{\given}{\, | \,}
\makeatletter
\@ifpackageloaded{caption}{}{\usepackage{caption}}
\AtBeginDocument{%
\ifdefined\contentsname
  \renewcommand*\contentsname{Table of contents}
\else
  \newcommand\contentsname{Table of contents}
\fi
\ifdefined\listfigurename
  \renewcommand*\listfigurename{List of Figures}
\else
  \newcommand\listfigurename{List of Figures}
\fi
\ifdefined\listtablename
  \renewcommand*\listtablename{List of Tables}
\else
  \newcommand\listtablename{List of Tables}
\fi
\ifdefined\figurename
  \renewcommand*\figurename{Figure}
\else
  \newcommand\figurename{Figure}
\fi
\ifdefined\tablename
  \renewcommand*\tablename{Table}
\else
  \newcommand\tablename{Table}
\fi
}
\@ifpackageloaded{float}{}{\usepackage{float}}
\floatstyle{ruled}
\@ifundefined{c@chapter}{\newfloat{codelisting}{h}{lop}}{\newfloat{codelisting}{h}{lop}[chapter]}
\floatname{codelisting}{Listing}

\makeatother
\makeatletter
\usepackage{pdflscape}
\makeatother
\makeatletter
\@ifpackageloaded{caption}{}{\usepackage{caption}}
\@ifpackageloaded{subcaption}{}{\usepackage{subcaption}}
\makeatother

\makeatletter
\xpatchcmd{\appendix}
  {\par}
  {\addcontentsline{toc}{section}{\@currentlabelname}\par}
  {}{}
\makeatother

\usepackage{etoolbox}

\makeatletter
\patchcmd{\LT@caption}
  {\bgroup}
  {\bgroup\global\LTpatch@captiontrue}
  {}{}
\patchcmd{\longtable}
  {\par}
  {\par\global\LTpatch@captionfalse}
  {}{}
\apptocmd{\endlongtable}
  {\ifLTpatch@caption\else\addtocounter{table}{-1}\fi}
  {}{}
\newif\ifLTpatch@caption
\makeatother

\begin{document}

\maketitle

\setcounter{secnumdepth}{5}

\setlength\LTleft{0pt}

Across the empirical sciences, the existence of an effect is commonly
inferred using frequentist null-hypothesis significance testing. Its key
outcome is the \(p\)-value which may be considered an absolute measure
of evidence against \(\mathcal{H}_0\) (\citeproc{ref-fisher1960}{Fisher,
1960}; \citeproc{ref-Wasserman2004}{Wasserman, 2004}), or an index of
\emph{surprise} (\citeproc{ref-Rafi2020}{Rafi \& Greenland, 2020}). The
\(p\)-value quantifies the conflict between the null hypothesis
\(\mathcal{H}_0\) and the observed data \(\mathbf{y}\), and consequently
the \(p\)-value is an asymmetric measure of evidence
(\citeproc{ref-Goodman1993}{Goodman, 1993};
\citeproc{ref-Goodman_1988}{Goodman \& Royall, 1988};
\citeproc{ref-Hubbard2008}{Hubbard \& Lindsay, 2008};
\citeproc{ref-Royall_2017}{Royall, 2017};
\citeproc{ref-Wasserstein2016}{Wasserstein \& Lazar, 2016}); for
instance, a small \(p\)-value does not necessarily indicate evidence in
favor of an alternative hypothesis \(\mathcal{H}_1\), and a large
\(p\)-value does not imply evidence in favor of \(\mathcal{H}_0\). In
contrast, from a Bayesian perspective, evidence is inherently relative.
A popular Bayesian measure of evidence is the ratio of the prior
predictive accuracy of one hypothesis relative to another
(\citeproc{ref-Edwards1963}{Edwards et al., 1963};
\citeproc{ref-Etz2017b}{Etz \& Wagenmakers, 2017};
\citeproc{ref-Jeffreys1935}{Jeffreys, 1935};
\citeproc{ref-Kass1995}{Kass \& Raftery, 1995}). This ratio is called
the Bayes factor \(\text{BF}_{01}\) and quantifies how much better
\(\mathcal{H}_0\) predicted the data compared to \(\mathcal{H}_1\),
\(\text{BF}_{01} = \text{p}(\mathbf{y} \mid \mathcal{H}_0) \; / \; \text{p}(\mathbf{y} \mid  \mathcal{H}_1)\).
The Bayes factor can also be interpreted as the change in a rational
observer's belief in \(\mathcal{H}_0\) relative to \(\mathcal{H}_1\)
brought about by the data. The relationship between the \(p\)-value and
the Bayes factor has been studied extensively, resulting in several
transformations from the \(p\)-value to \emph{bounds} on the Bayes
factor (e.g., \citeproc{ref-Edwards1963}{Edwards et al., 1963};
\citeproc{ref-Held2018}{Held \& Ott, 2018}). For instance, Sellke et al.
(\citeproc{ref-Sellke_2001}{2001}) proposed that when \(p<1/e\), the
Bayes factor in favor of \(\mathcal{H}_1\) is no greater than
\(1/(-e p \log{p})\); with \(p = .05\) this yields an upper bound of
about \(2.46\). In other words, when \(p = .05\) the data are not even
2.5 times more likely under \(\mathcal{H}_1\), across a wide class of
prior distributions, than under \(\mathcal{H}_0\). When \(p = .005\)
(\citeproc{ref-Benjamin_2017}{Benjamin et al., 2017}) the upper bound is
\(13.89\).

Here we aim to clarify the relationship between the \(p\)-value and the
Bayes factor for objective Bayesian hypothesis testing of the existence
of an effect (i.e., when prior distributions under \(\mathcal{H}_1\) are
relatively uninformed and selected based on general desiderata and
\(\mathcal{H}_0\) is a point-null hypothesis;
\citeproc{ref-Consonni2018}{Consonni et al., 2018};
\citeproc{ref-Bayarri2012}{Bayarri et al., 2012};
\citeproc{ref-Liang2008}{Liang et al., 2008}). Specifically, we show
that a surprisingly simple transformation of the \(p\)-value can yield
an accurate approximation of the objective Bayes factor. This
transformation, called Jeffreys's Approximate Bayes factor (JAB), dates
back to the work of Harold Jeffreys in the second half of the 1930s
(e.g., \citeproc{ref-Jeffreys1935}{Jeffreys, 1935},
\citeproc{ref-Jeffreys1939}{1939}; see also
\citeproc{ref-Wagenmakers_2022}{Wagenmakers \& Ly, 2022}) but appears to
have gone largely unnoticed until recently
(\citeproc{ref-Velidi2025}{Velidi et al., 2025};
\citeproc{ref-Wagenmakers2022}{Wagenmakers, 2022}). In its simplest
form, Jeffreys's approximation requires only the \(p\)-value and the
effective sample size \(n_\text{eff}\)
(\citeproc{ref-Wagenmakers2022}{Wagenmakers, 2022}): Assuming a
unit-information prior for the test-relevant parameter under
\(\mathcal{H}_1\), the approximate Bayes factor in favor of
\(\mathcal{H}_0\) is \(\text{JAB}_{01} \approx 3p \sqrt{n_\text{eff}}\)
when \(p < .10\). For example, when \(p = .05\) and \(n = 100\),
\(\text{JAB}_{01} = 1.5\), which means that the data are actually 1.5
times more likely under \(\mathcal{H}_0\) than under \(\mathcal{H}_1\);
when \(p = .005\) and \(n = 100\), \(\text{JAB}_{01} \approx 1/6.7\),
which means that the data are 6.7 times more likely under
\(\mathcal{H}_1\) than under \(\mathcal{H}_0\). JAB therefore bridges
two seemingly incompatible concepts: Bayesian change in belief and
frequentist surprise. In contrast to earlier work, JAB does not involve
an upper or lower bound---JAB involves a direct mapping between the
\(p\)-value and the objective Bayes factor. As will become apparent
later, this requires that JAB includes a \(\sqrt{n_\text{eff}}\) term.
We focus here on tests of existence. In tests of direction a simple
(approximate) relationship between \(p\)-value and the Bayes factor has
long been known: When assuming the location in a model from the
exponential family is \(\mu \neq 0\) and the goal is to infer its sign
(positive vs. negative, \(\mathcal{H}_+: \theta > 0\) and
\(\mathcal{H}_-: \theta < 0\)) the one-sided \(p\)-value equals the
posterior probability that the effect is negative (with a prior
symmetric around 0; \citeproc{ref-Jeffreys1961}{Jeffreys, 1961, p.
387}). Transforming the \(p\)-value to odds yields an approximate Bayes
factor, for example when assuming equal prior odds,
\(\text{BF}_{-+} = p / (1 - p)\) (\citeproc{ref-Casella_1987}{Casella \&
Berger, 1987}; \citeproc{ref-Marsman2016}{Marsman \& Wagenmakers,
2016}).

The remainder of this manuscript is organized as follows.
Section~\ref{sec-conflict-vs-relative-support} briefly reviews the
evidential interpretation of the \(p\)-value as a measure of conflict or
surprise and illustrates why, in contrast to the Bayes factor, a
\(p\)-value cannot quantify support in favor of \(\mathcal{H}_0\). In
Section~\ref{sec-jab}, we introduce JAB and how it combines prior
information, effective sample size, and the \(p\)-value. We review the
assumptions underlying the approximation to clarify when it is
applicable and when it may fail. We then show that, in practice, JAB is
a fair approximation to commonly used default Bayes factors for
comparisons of means and proportions. In
Section~\ref{sec-evidence-in-p}, we discuss the implications of JAB for
previously suggested evidential interpretations of ranges of \(p\)
values.

\section{Evidence as conflict or as relative
support}\label{sec-conflict-vs-relative-support}

The one-tailed \(p\)-value is the percentile of the observed test
statistic \(t\) in the sampling distribution of \(T\) under the null
hypothesis \(\mathcal{H}_0\),

\begin{equation}\protect\phantomsection\label{eq-p}{
\begin{aligned}
p & = \text{Pr}(T \geq t \mid \mathcal{H}_0) \text{ or} \\
p & = \text{Pr}(T \leq t \mid \mathcal{H}_0),    
\end{aligned}
}\end{equation} depending on the hypothesized direction of the effect.
That is, \(p\) is the probability of obtaining data that contradict
\(\mathcal{H}_0\) at least as strongly as the observed data, if
\(\mathcal{H}_0\) were true. As we are concerned with measuring
statistical evidence, we follow a (neo-)Fisherian interpretation of the
\(p\)-value (\citeproc{ref-Christensen_2005}{Christensen, 2005}).
According to this view, the \(p\)-value quantifies conflict between the
data and \(\mathcal{H}_0\) (\citeproc{ref-Greenland2019}{Greenland,
2019}; \citeproc{ref-Perezgonzalez2015}{Perezgonzalez, 2015}) and can be
expressed as \emph{surprise}, \(s = -\log_2(p)\), in units of bits (or
Shannon-information; \citeproc{ref-Rafi2020}{Rafi \& Greenland, 2020}).
As noted earlier, evidence is defined differently in the Bayesian
paradigm. In the following we refer to the evidence concept underlying
\(p\)-values as ``surprise'' and reserve ``evidence'' for the Bayes
factor---as is customary in Bayesian terminology. To appreciate the
difference between surprise and evidence, consider a series of five coin
tosses. Assuming fair coin tossing, a streak of all heads from five
tosses roughly corresponds to \(p = .031 = .5^{5}\) in a one-sided exact
binomial test, which translates to a surprise of \(s = 5\)
(\citeproc{ref-Greenland2019}{Greenland, 2019, p. 109}). How should this
surprise change our belief? This depends on the nature of the
alternative explanations. As is clear from Equation~\ref{eq-p}, the
\(p\)-value does not consider alternatives to \(\mathcal{H}_0\)---the
fair coin. Hence, the relationship between surprise and belief is
unclear.

In the following we attempt to convey an intuitive understanding of the
Bayes factor and how it differs from the \(p\)-value. For a more
in-depth treatment see Al-Labadi et al.
(\citeproc{ref-Allabadi2024}{2025}) and Morey et al.
(\citeproc{ref-Morey2016}{2016}). According to Bayes' theorem, evidence
is the change in relative belief brought about by data and referred to
as the Bayes factor. Specifically, the Bayes factor \(\text{BF}_{0i}\)
is the updating factor that reflects the change from prior odds of
\(\mathcal{H}_0\) relative to \(\mathcal{H}_i\) to posterior odds,

\[
\underbrace{ \frac{\text{Pr}(\mathcal{H}_0  \mid \mathbf{y})}{\text{Pr}(\mathcal{H}_i  \mid \mathbf{y})}}_{\substack{\text{Posterior beliefs}\\ \text{about hypotheses}} } \,\, =
\underbrace{ \frac{\text{Pr}(\mathcal{H}_0)}{\text{Pr}(\mathcal{H}_i)}}_{\substack{\text{Prior beliefs}\\ \text{about hypotheses}} }
\times \,\,\,\,
\underbrace{ \frac{\text{p}(\mathbf{y} \mid \mathcal{H}_0)}{\text{p}(\mathbf{y} \mid  \mathcal{H}_i)}}_{\substack{\text{Bayes factor BF}_{0i}\\ (\text{relative evidence}) } }.
\]

This expression makes clear that evidence is inherently relative;
without an alternative hypothesis, we can quantify hypothesis-data
conflict (Equation~\ref{eq-p}), but never the change in belief brought
about by the data.

For our coin tossing example, in addition to good fortune, we consider
two alternative explanations for a series of 5 heads out of 5 tosses:
First, with any game of chance there is always the danger of deceit, so
we entertain the possibility that the coin shows heads on both sides,
\(\mathcal{H}_1: \theta = 1\). Second, maybe the coin tosser always
started with heads up, thereby inflating the probability of heads.
Informed by the empirical estimate reported by
(\citeproc{ref-Bartos2024}{Bartoš et al., 2025}) we represent this
hypothesis by a Beta distribution with mean \(\mu_\theta = 0.515\) and
standard deviation \(\sigma_\theta = 0.008\),
\(\mathcal{H}_2: \theta \sim \mathcal{B}(a = 1910, b = 1800)\).

What is the relative evidence that the coin toss was fair, that heads
and tails are equally likely? As noted above, a run of five heads
corresponds to \(p = .031 = 0.5^5\) under \(\mathcal{H}_0\).
Conveniently, in this example
\(\Pr(\mathbf{y} \mid \mathcal{H}_0) = \theta^5 = p\)---this is
typically not the case. If \(\mathcal{H}_1\) is true,
\(\Pr(\mathbf{y} \mid \mathcal{H}_1) = 1^5 = 1\). Therefore, a run of 5
heads gives \(\text{BF}_{01} = p/1 = .031 = 1/32\)---it is 32 times more
likely under \(\mathcal{H}_1\) than under \(\mathcal{H}_0\). The data
provide \emph{strong} evidence that the coin shows heads on both sides
(\citeproc{ref-Lee2014}{Lee \& Wagenmakers, 2014};
\citeproc{ref-Wagenmakers2024}{Wagenmakers \& Aust, 2024}). Counter to
our initial claims, in this example
\(p = 1/\text{BF}_{10} = \text{BF}_{01}\) quantifies evidence for
\(\mathcal{H}_0\) relative to \(\mathcal{H}_1\). This is an unlikely
coincidence that is useful for a brief exposition but this equivalence
immediately breaks down for different data or other alternative
hypotheses. Had the outcome of the fifth flip been tails---not
heads---we would see \(p = .187\) and should be less surprised. But now
\(p \neq\Pr(\mathbf{y} \mid \mathcal{H}_0) = .156\) and there is a stark
difference between \(p\) and the Bayes factor:
\(\Pr(\mathbf{y} \mid \mathcal{H}_1) = 0\) and hence
\(\text{BF}_{01} = \infty\). Whereas the \(p\)-value rightly indicates
modest conflict between the data and \(\mathcal{H}_0\), the Bayes factor
shows definitive \emph{support} for
\(\mathcal{H}_0\)---\(\mathcal{H}_1\) has been discredited beyond
repair. This is not reflected in the \(p\)-value because it is not
concerned with \(\mathcal{H}_1\). See this conflict would require
calculating another \(p\) for \(\mathcal{H}_1\). Returning to the
original example of 5 heads out of 5 tosses, the equivalence between
\(p\) and the Bayes factor also breaks down for any alternative
hypothesis other than \(\mathcal{H}_1\): Relative to the biased coin
tosser (\(\mathcal{H}_2\)) the data provide weak evidence against
\(\mathcal{H}_0\), \(\text{BF}_{02} = 1/1.17\). A run of 5 heads is 1.17
times more likely if the coin was tossed with a slight heads-bias than
if it was tossed fairly. The data are essentially uninformative. The
evidence in this case is weaker because compared to \(\mathcal{H}_1\)
the predictions of \(\mathcal{H}_2\) are much more similar to those of
\(\mathcal{H}_0\). Taken together, this example shows that there is no
direct relationship between Bayesian evidence and frequentist surprise.

\section{Jeffreys's Approximate Bayes Factor}\label{sec-jab}

As we discussed, the \(p\)-value measures surprise under
\(\mathcal{H}_0\), whereas the Bayes factor measures evidence for or
against \(\mathcal{H}_0\) relative to \(\mathcal{H}_1\). Against this
backdrop, it may be unexpected that a simple transformation of the
\(p\)-value yields a remarkably good approximation to the Bayes factor.
We briefly explain and illustrate this approximation known as Jeffreys's
Approximate Bayes factor (JAB) and then explore how it relates to the
\(p\)-value and surprise.

\begin{figure}

\caption{\label{fig-jabp-jzs}Linear relationships between JZS-Bayes
factor and \(p\)-value-based JAB for 704 \(t\)-test results collected by
Aczel et al. (\citeproc{ref-Aczel_2018}{2018}) and Wetzels et al.
(\citeproc{ref-Wetzels_2011}{2011}). Triangles represent one-sided
\(p\)-values, circles represent \(\text{JAB}_{01}\), each on logarithmic
scale. The color of points indicates the effective sample size. The
solid grey line shows the estimated linear relationship between
\(p\)-values and Bayes factors. The grey area shows the margin of error
of a factor of 3 (\citeproc{ref-Jeffreys1961}{Jeffreys, 1961, p. 433}).}

\centering{

\pandocbounded{\includegraphics[keepaspectratio]{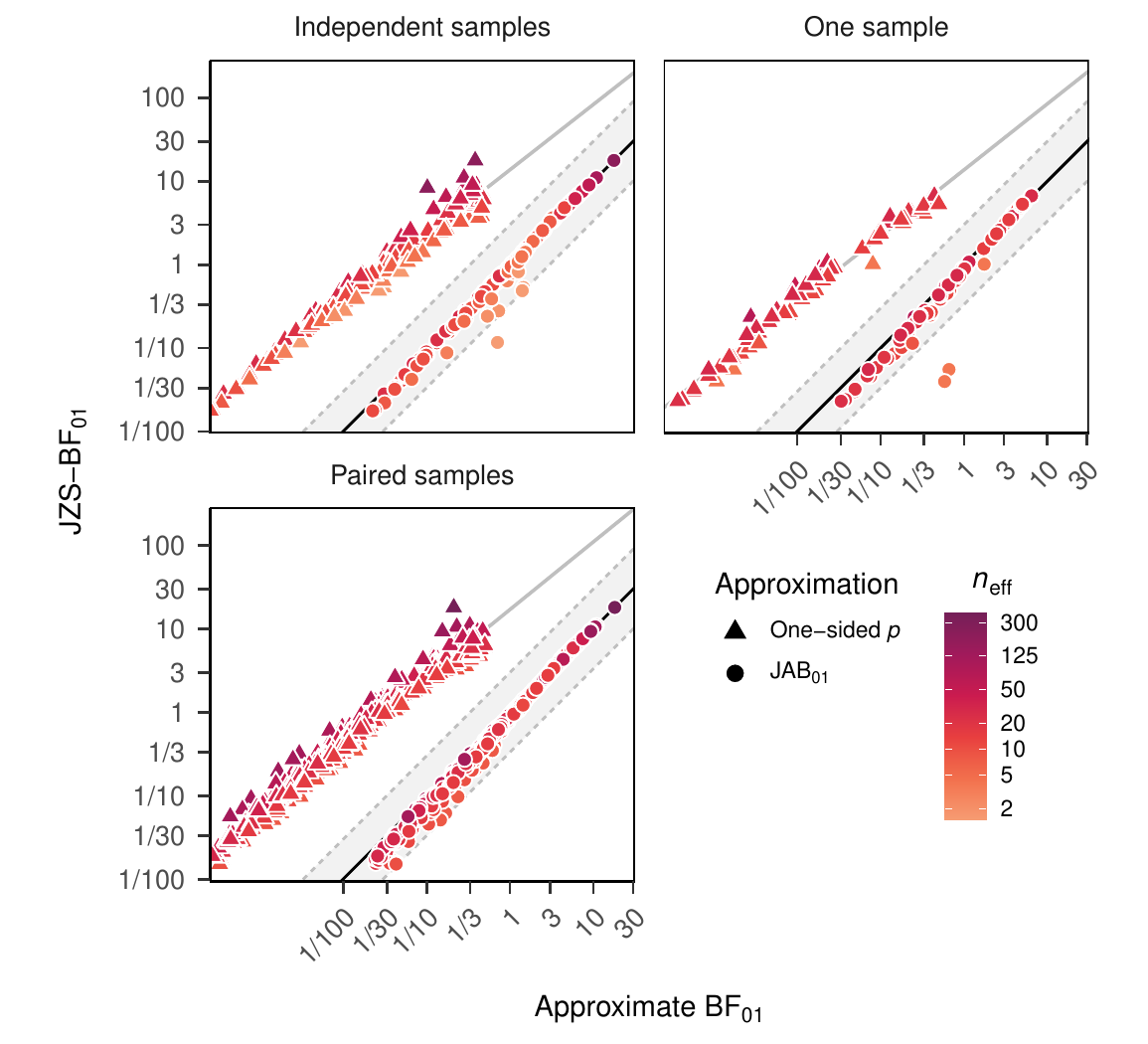}}

}

\end{figure}

Berger and Sellke (\citeproc{ref-Berger1987}{1987}) showed that there is
a monotonic relationship between \(p\) and the lower bound of the Bayes
factor for \(\mathcal{H}_0\). For a given sample size, larger effects
yield smaller \(p\)-values and stronger evidence against
\(\mathcal{H}_0\). Marsman and Wagenmakers
(\citeproc{ref-Marsman2016}{2016}) showed that, for location parameters
\(\mu\) in the exponential family (e.g., a normal distribution) and a
given sample size \(n\), the logarithm of the one-sided \(p\)-value is
approximately linearly related to the Bayes factor. We illustrate this
relationship in Figure~\ref{fig-jabp-jzs} for the published \(t\)-test
collected by Aczel et al. (\citeproc{ref-Aczel_2018}{2018}) and Wetzels
et al.~(\citeproc{ref-Wetzels_2011}{2011}; reanalyzed by
\citeproc{ref-Rouder_2012}{Rouder et al., 2012}). Triangles show the
linear relationship to the commonly used JZS-Bayes factor with a
standard Cauchy prior (\citeproc{ref-Rouder_2009}{Rouder et al., 2009})
on logarithmic scale. Two regularities are worth noting: (1) The
logarithm of the one-sided \(p\)-values is substantially smaller than
the Bayes factor---the surprise is considerably larger than the relative
evidence against \(\mathcal{H}_0\). (2) There is systematic variability
around the best fitting line: \(p\)-values for large samples fall above
the line, while \(p\)-values for small samples fall below the line.
Hence, despite being linearly related to the Bayes factor, surprise is
not directly related to relative evidence.

Wagenmakers (\citeproc{ref-Wagenmakers2022}{2022}) recently highlighted
that \(p\)-values of single-parameter Wald tests are directly related to
JAB (\citeproc{ref-Jeffreys1935}{Jeffreys, 1935}; see
\citeproc{ref-Velidi2025}{Velidi et al., 2025} for a generalization).
One of the main insights from Jeffreys's work was that an approximate
Bayes factor can be obtained from the Wald statistic

\[
W = \left( \frac{\hat \theta - \theta_0}{\text{SE}(\hat \theta)} \right)^2
\] for a test-relevant parameter \(\theta\) and the null value
\(\theta_0\) as

\begin{equation}\protect\phantomsection\label{eq-jab}{
\begin{aligned}
\text{JAB}_{01} & = \frac{1}{\sqrt{2 \pi}~\text{SE}(\hat \theta)~g(\hat \theta \given \mathcal{H}_1)}  \; \exp(-0.5 \; W), & \text{with } \text{SE}(\hat\theta) = \sigma/\sqrt{n_\text{eff}} \\
& = A \; \sqrt{n_\text{eff}} \; \exp(-0.5 \; W), &
\end{aligned}
}\end{equation} where
\(A = [\sqrt{2 \pi}~\sigma~g(\hat \theta \given \mathcal{H}_1)]^{-1}\)
depends on the prior distribution \(g()\) under \(\mathcal{H}_1\)
evaluated at the maximum likelihood estimate of the test-relevant
parameter \(\hat \theta\), \(\sqrt{n_\text{eff}}\) is the
\emph{effective sample size}, and \(\sigma\) denotes the asymptotic
standard deviation coefficient (see Appendix~Section~\ref{sec-neff}). In
models where \(n_\text{eff}\) corresponds to an information-equivalent
sample size, \(\sigma^2\) is the variance associated with one unit of
information. JAB relates to two-sided \(p\)-values through the quantile
function \(F^{-1}()\) of the asymptotic sampling distributions of the
corresponding Wald test statistic---the
\(\chi^2(\mathrm{df} = 1)\)-distribution or the standard normal
distribution \(\mathcal{N}(\mu = 0, \sigma^2 = 1)\),

\begin{equation}\protect\phantomsection\label{eq-w-p}{
\begin{aligned}
W & = \phantom{[} F^{-1}_{\chi^2(1)}(1-p) \phantom{]^2} & \text{for } \chi^2\text{-tests and} \\
  & = \left\{ F^{-1}_{\mathcal{N}(0,1)}(p/2) \right\}^2 & \text{for } z\text{-tests.}
\end{aligned}
}\end{equation} In words, the latter is the square of the
probit-transformed one-sided \(p\)-value. So, JAB can be understood as a
transformation of the Wald \(p\)-value, using the sample size, to
relative Bayesian evidence. The circles in Figure~\ref{fig-jabp-jzs}
show that, with the same standard Cauchy prior (scale \(r = 1\)), JAB
approximates the JZS-Bayes factor quite well. In contrast to the
one-sided \(p\)-value, JAB largely accounts for differences in evidence
related to effective sample size. Because these \(p\) values are from a
\(t\) test---rather than a Wald test---JAB understates the evidence for
\(\mathcal{H}_1\) when \(n_\text{eff}\) is small and the evidence in
favor of \(\mathcal{H}_1\) is strong. However, as pointed out by
Jeffreys (\citeproc{ref-Jeffreys1961}{1961}),

\begin{quote}
We do not need {[}the Bayes factor{]} with much accuracy. (\ldots) it
will seldom matter appreciably to further procedure if {[}the Bayes
factor{]} is wrong by as much as a factor of 3. (pp.~432-433)
\end{quote}

For the real-world data used here, the bias exceeds a factor of 3 only
in very small samples. This suggests that in most situations, the
\(p\)-based JAB is a fair approximation to the JZS-Bayes factor for
\(t\)-tests. Under standard regularity conditions, \(W\) asymptotically
converges to \(\Lambda = 2 \; \log(\mathcal{L}_{1} / \mathcal{L}_{0})\),
with \(\mathcal{L}_{i}\) the likelihood of model \(i\)
(\citeproc{ref-Buse_1982}{Buse, 1982}; \citeproc{ref-Engle1984}{Engle,
1984, p. 798}; \citeproc{ref-Vaart_1998}{Vaart, 1998, p. 227}). When the
exact likelihood ratio is used instead in Equation~\ref{eq-jab}, JAB
approximates the JZS Bayes factor even more closely. Deviations are
noticeable only in very small samples, see Appendix~\ref{sec-llr}.

As noted above, the linear relationship between the logarithms of the
one-sided \(p\)-value and the Bayes factor shown in
Figure~\ref{fig-jabp-jzs} only holds for tests of location parameters in
the exponential family (e.g., assuming normally distributed errors).
When the tested hypothesis is of a different kind, this relationship can
break down. Consider the example of comparing two independent
proportions \(\theta_1\) and \(\theta_2\), where
\(\mathcal{H}_0: \theta_1 - \theta_2 = 0\). When the number of
observations is large and the probabilities are not too extreme, we can
use the \(z\)-test for two proportions (or equivalently, Pearson's
\(\chi^2\)-test for 2 \(\times\) 2-contingency tables). Another option
is to reformulate the hypothesis in terms of the log odds ratio
\(\log(\text{OR})\),

\[
\mathcal{H}_0: \log(\text{OR}) = \log\left\{\frac{\theta_1/(1-\theta_1)}{\theta_2/(1-\theta_2)}\right\} = 0,
\] and use a logistic regression model with an asymptotic \(z\)-test.
Corresponding Bayesian hypothesis tests are available
(\citeproc{ref-Dablander_2021}{Dablander et al., 2021};
\citeproc{ref-Howard_1998}{Howard, 1998}).

\begin{figure}

\caption{\label{fig-jab-prop}Linear relationships between Bayes factors
and \(p\)-value-based JAB for 39 results of comparisons of two
proportions collected by Hoekstra et al.
(\citeproc{ref-Hoekstra_2018}{2018}; reanalyzed by
\citeproc{ref-Dablander_2021}{Dablander et al., 2021}). The top panel
shows results for \(z\)-tests and its Bayesian analog using independent
Beta-priors (IB); the bottom panel shows results from logistic
regression analysis and its Bayesian analog (LT). Triangles represent
one-sided \(p\)-values, circles represent \(\text{JAB}_{01}\), each on
logarithmic scale. The color of points indicates the effective sample
size. The grey area shows the margin of error of a factor of 3
(\citeproc{ref-Jeffreys1961}{Jeffreys, 1961, p. 433}).}

\centering{

\pandocbounded{\includegraphics[keepaspectratio]{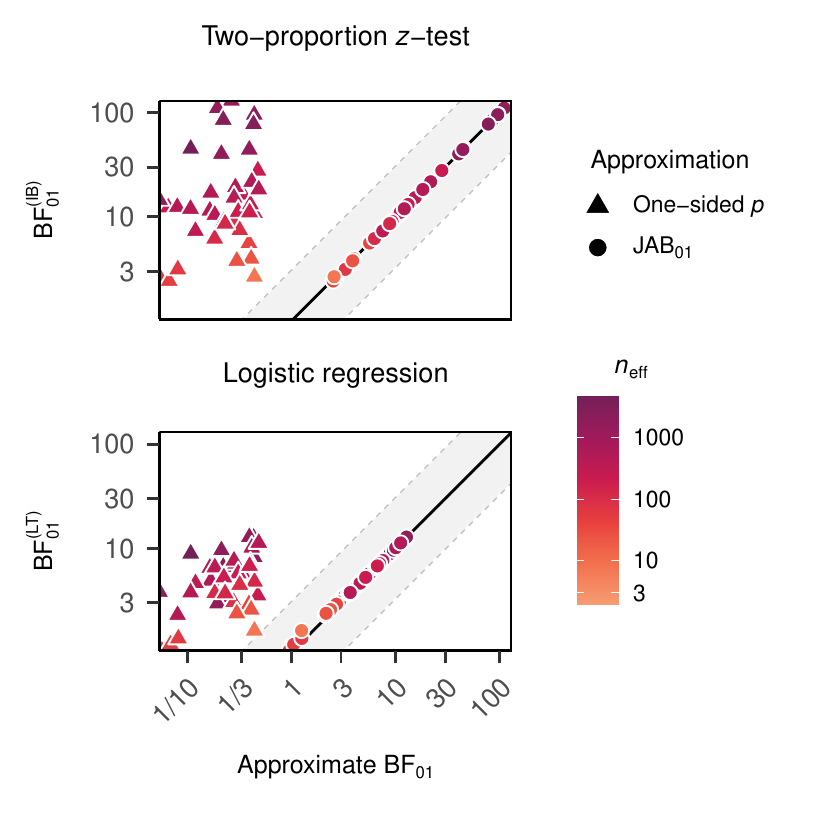}}

}

\end{figure}

In Figure~\ref{fig-jab-prop}, we plot the results of 39 published
comparisons of two independent proportions collected by Hoekstra et al.
(\citeproc{ref-Hoekstra_2018}{2018}; reanalyzed by
\citeproc{ref-Dablander_2021}{Dablander et al., 2021}). For the Bayesian
analog to the two-proportion \(z\)-test we place independent uniform
Beta-priors (IB) on the two proportions, which implies a symmetric,
zero-centered triangular prior on the difference
\(\theta_1 - \theta_2 \in [-1, 1]\); for the logistic regression
analysis we use a standard normal distribution as prior for
\(\log(\text{OR})\) (\citeproc{ref-Dablander_2021}{Dablander et al.,
2021}). We used the same priors for JAB. For neither hypothesis tests
does the one-sided \(p\)-value exhibit a clear relationship to the
corresponding Bayes factor. JAB, on the other hand, is again closely
related to the corresponding Bayes factors.

These two examples suggest that JAB can provide a generic, simple, and
accurate approximation to objective Bayes factors---a remarkable result
because it shows that by combining \(p\)-values with sample size one may
obtain an objective Bayes factor, even though \(p\)-values and Bayes
factors have very different statistical properties and philosophical
foundations.

\subsection{Assumptions}\label{assumptions}

Before we use JAB to explore how surprise relates to evidence, we wish
to highlight the assumptions underlying JAB to clarify the boundary
conditions of the approximation. JAB approximates the Bayes factor
assuming that the sampling distribution of \(\hat\theta\) is
(asymptotically) normal,

\[
\text{BF}_{01} = \frac{\text{p}(\hat\theta \given \mathcal{H}_0)}{\text{p}(\hat\theta \given \mathcal{H}_1)} = \frac{\mathcal{N}(\hat\theta \given \theta_0, \text{SE}(\hat\theta)^2)}{\int \mathcal{N}(\hat\theta \given \theta, \text{SE}(\hat\theta)^2) \; g(\theta \given \mathcal{H}_1) \; \text{d}\theta} \approx \frac{\mathcal{N}(\hat\theta \given \theta_0, \text{SE}(\hat\theta)^2)}{g(\hat\theta \given \mathcal{H}_1)} = \text{JAB}_{01}.
\] While the density of \(\hat\theta\) under \(\mathcal{H}_0\),
\(\mathcal{N}(\hat\theta \given \theta_0, \text{SE}(\hat\theta)^2)\), is
known, the integral in the denominator is not readily available. It is
assumed that this integral can be approximated by the prior density of
the maximum likelihood estimate under \(\mathcal{H}_1\),
\(g(\hat\theta \given \mathcal{H}_1)\)
(\citeproc{ref-Jeffreys1961}{Jeffreys, 1961, p. 247} ff.). This
assumption results from using a Taylor expansion to solve the integral
(Appendix \ref{sec-jab-derivation}). Hence, the following assumptions
must hold for JAB to be accurate:

\textbf{Assumption 1}. Either \(\theta\) is orthogonal to all nuisance
parameters shared by \(\mathcal{H}_0\) and \(\mathcal{H}_1\) or the
priors on the nuisance parameters are vague
(\citeproc{ref-Jeffreys1961}{Jeffreys, 1961, pp. 249--251}). If
\(\theta\) is a function of \emph{other} model parameters with informed
priors, JAB can be severely biased.

\textbf{Assumption 2}. The sampling distribution of the maximum
likelihood estimate is normally distributed,
\(\hat\theta \sim \mathcal{N}(\theta, \text{SE}(\hat\theta)^2)\). In
many cases, this assumption holds asymptotically but it can be violated
when the sampling distribution of the raw data from which \(\hat\theta\)
is computed is not normal and the sample size is small (see also
Appendix~\ref{sec-llr}).

\textbf{Assumption 3}. The prior distribution is twice continuously
differentiable, proper, and assigns positive probability density to
\(\hat\theta\). If \(g(\hat\theta \given \mathcal{H}_1) = 0\),
\(\text{JAB}_{01}\) is undefined.

\begin{figure}

\caption{\label{fig-curvature-prior}Curvature of prior distributions as
a funciton of \(\theta\) for different prior scales.}

\centering{

\pandocbounded{\includegraphics[keepaspectratio]{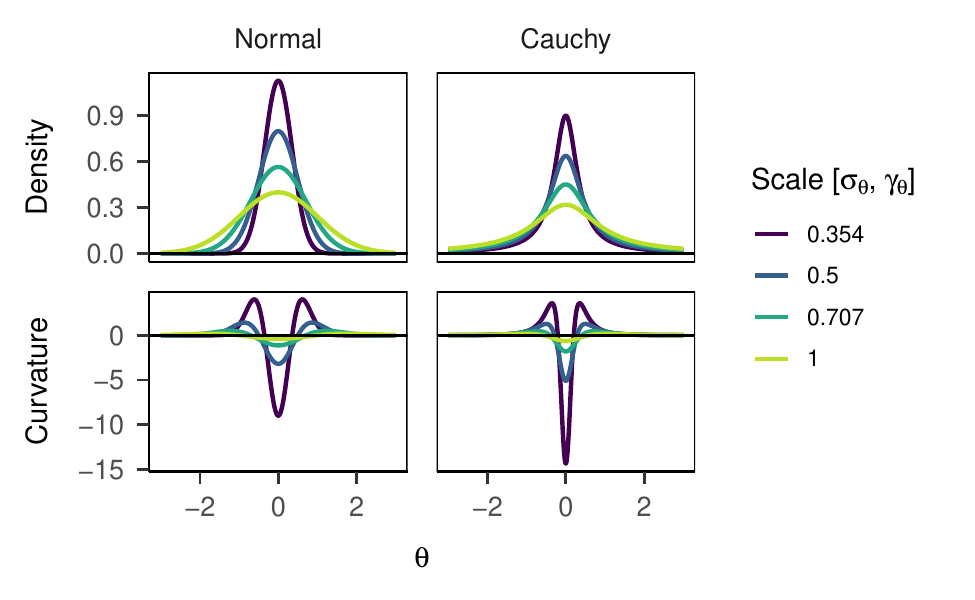}}

}

\end{figure}

\textbf{Assumption 4}. The standard error is small or the prior
distribution at \(\hat\theta\) is only mildly curved (concave or convex,
Appendix~\ref{sec-jab-derivation}. The standard error, of course,
decreases as the effective sample size \(n_\text{eff}\) increases. As
illustrated in Figure~\ref{fig-curvature-prior}, the curvature of the
prior distribution depends on the distribution family but in
location-scale distributions it typically decreases as its scale
increases. Hence, the accuracy of the approximation is higher for
relatively wide, uninformative priors. Broadly speaking, the data must
be informative relative to the prior distribution,
\(\text{SE}(\hat\theta) \ll \sigma_\theta\) (Appendix
\ref{sec-jab-derivation}), or JAB will be biased against
\(\mathcal{H}_0\) when \(\hat\theta\) is near the peak and biased
against \(\mathcal{H}_1\) when \(\hat\theta\) is in the tail of the
prior distribution.

\textbf{Assumption 5}. When \(W\) is calculated from \(p\)-values
(Equation~\ref{eq-w-p}), it is further assumed that the \(p\)-value was
not corrected for multiple comparisons, as otherwise the quantile
back-transformation of the \(p\)-value to \(W\) is incorrect.

As a rule of thumb, these assumptions likely hold in large samples, or
for normally distributed \(\hat\theta\) if the prior distribution is
relatively wide, uninformed and places non-negligible mass near
\(\hat\theta\). That is, in most applications JAB will be accurate for
objective hypothesis testing with relatively uninformed prior
distributions. This explains why JAB is remarkably accurate in our
applications (i.e., Figure~\ref{fig-jabp-jzs} and
Figure~\ref{fig-jab-prop}).

\subsection{Accuracy of the approximation}\label{sec-jab-accuracy}

To assess the quality of the approximation, we compare
\(\text{JAB}_{01}\) to the analytic Bayes factor using a normal prior
distribution. Substituting the normal prior
\(g(\hat \theta \mid \mathcal{H}_1) = (2\pi\sigma_\theta^2)^{-1/2} \exp\left\{-0.5 [(\hat\theta - \mu_\theta)/\sigma_\theta]^2\right\}\)
into Equation~\ref{eq-jab} yields

\begin{equation}\protect\phantomsection\label{eq-jab-normal}{
\text{JAB}_{01}
  = \sqrt{\frac{\sigma_\theta^2}{\text{SE}(\hat\theta)^2}}
    \;
    \left\{ \exp\left(-\frac{1}{2}\,\frac{(\hat\theta - \mu_\theta)^2}{\sigma_\theta^2}\right) \right\}^{-1}
    \;
    \exp\left(-\frac{1}{2}\,W\right).
}\end{equation} For the same prior, the analytic \(\text{BF}_{01}\)
(\citeproc{ref-Barto__2023}{Bartoš \& Wagenmakers, 2023};
\citeproc{ref-Berger_1987}{Berger \& Delampady, 1987}) is

\begin{equation}\protect\phantomsection\label{eq-analytic-bf}{
\text{BF}_{01}
  = \sqrt{\frac{\sigma_\theta^2 + \text{SE}(\hat\theta)^2}{\text{SE}(\hat\theta)^2}}
    \;
    \left\{ \exp\left(-\frac{1}{2}\,\frac{(\hat\theta - \mu_\theta)^2}{\text{SE}(\hat\theta)^2 + \sigma_\theta^2}\right) \right\}^{-1}
    \;
    \exp\left(-\frac{1}{2}\,W\right).
}\end{equation} Both expressions share the same likelihood ratio
component with the Wald statistic \(W\). They differ in that JAB
approximates the marginal likelihood by evaluating the prior at a single
point (\(\hat\theta\)), whereas the analytic Bayes factor accounts for
the uncertainty in \(\hat\theta\) adding to the prior variance
(\(\sigma_\theta^2 + \text{SE}^2\)). Hence, for a fixed \(\hat\theta\)
and \(\text{SE}(\hat\theta)\) the bias of \(\text{JAB}_{01}\) relative
to \(\text{BF}_{01}\) is entirely a function of the prior:

\begin{equation}\protect\phantomsection\label{eq-bias-bf}{
k = \frac{\text{BF}_{01}}{\text{JAB}_{01}}
  = \underbrace{\sqrt{\frac{\sigma_\theta^2 + \text{SE}(\hat\theta)^2}{\sigma_\theta^2}}}_{\text{variance inflation}}
    \;\exp\left\{
      -\frac{1}{2} \,
      \frac{(\hat\theta - \mu_\theta)^2}{\sigma_\theta^2}
      \,
      \underbrace{\frac{\text{SE}(\hat\theta)^2}{\text{SE}(\hat\theta)^2 + \sigma_\theta^2}}_{\text{overconfidence}}
    \right\}.
}\end{equation} This reveals two opposing biases: The factor
\(\sqrt{[\,\sigma_\theta^2 + \text{SE}(\hat\theta)^2\,] \,/\, \sigma_\theta^2}\)
represents a variance inflation factor that biases \(\text{JAB}_{01}\)
towards \(\mathcal{H}_1\) (\(k > 1\)), while the exponential term
reflects the overconfidence of evaluating the prior at the MLE, which
can bias JAB towards \(\mathcal{H}_0\) (\(k < 1\)) when \(\hat\theta\)
is far from \(\mu_\theta\). The resulting bias will be small if
\(\sigma_\theta^2 \gg \text{SE}(\hat{\theta})^2\) and \(\mu_\theta\) is
not too far from \(\hat{\theta}\) (in units of \(\sigma_\theta^2\)). In
objective Bayesian hypothesis testing, this is generally reasonable to
assume unless the sample size is small.

To make this more concrete, consider the following two variants of the
unit-information prior---a common choice in objective Bayesian
hypothesis testing. First, the unit-information prior is centered on the
test value \(\theta_0\) with a variance calibrated so that the prior
contains as much information as one observation
(\citeproc{ref-Kass1995}{Kass \& Raftery, 1995}). For the probability of
a coin toss, a unit-information prior to test if a coin is fair
(\(\theta_0 = 0.5\)) corresponds to a \(\mathcal{B}(a = 0.5, b = 0.5)\)
prior distribution over the probability of heads. Second, centering the
unit-information prior on the maximum likelihood estimate
\(\hat\theta\)---instead of the test value \(\theta_0\)---yields a
data-dependent pseudoprior that is mathematically convenient and
provides the theoretical justification to use the BIC as an
approximation of the Bayes factor (\citeproc{ref-Kass1995}{Kass \&
Raftery, 1995}; \citeproc{ref-Schwarz_1978}{Schwarz, 1978}) For this
prior, Equation Equation~\ref{eq-bias-bf} simplifies to

\begin{equation}\protect\phantomsection\label{eq-bias-bf-uip}{
k = \sqrt{1 + 1/n_\text{eff}}.
}\end{equation}

This has two implications: (1) This JAB can never overstate the evidence
in favor of the alternative by more than a factor of \(\sqrt{2} = 1.41\)
or about 40\%, and the bias quickly becomes negligible as
\(n_\text{eff}\) increases. (2) The bias of \(\text{JAB}_{01}\) is
equivalent to the impact of a single observation
(\citeproc{ref-Barto__2023}{Bartoš \& Wagenmakers, 2023}),

\begin{equation}\protect\phantomsection\label{eq-upi-jab-corrected}{
\text{BF}_{01} = \text{JAB}_{01} \; \sqrt{1 + 1/n_\text{eff}} = \sqrt{n_\text{eff} + 1} \; \exp\left( -0.5 \; W \right).
}\end{equation}

\begin{figure}

\caption{\label{fig-jab-bias-contour}Bias of \(\text{JAB}_{01}\) as a
function of standardized effect size \(\hat\delta\) and effective sample
size \(n_\text{eff}\). Solid colored lines show the exact bias, dashed
lines show the large-sample approximation. The solid black line shows no
bias for reference.}

\centering{

\pandocbounded{\includegraphics[keepaspectratio]{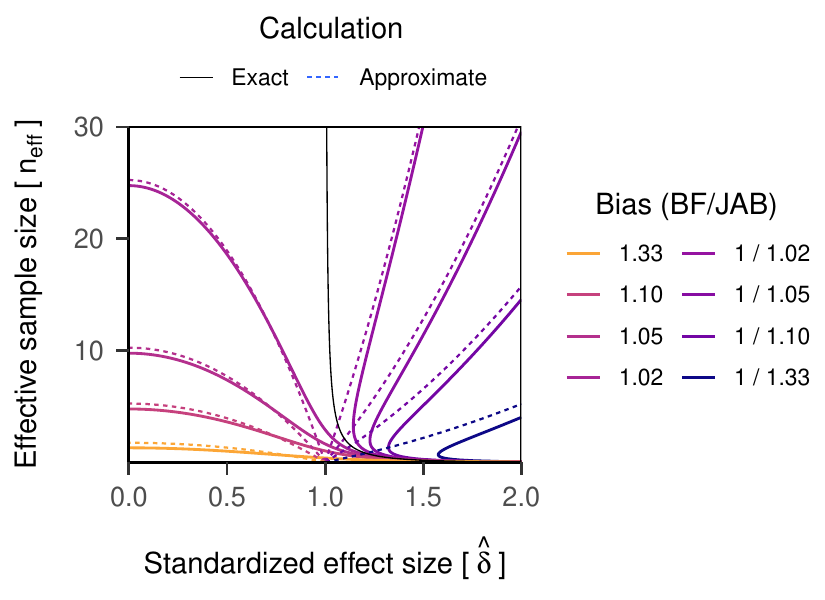}}

}

\end{figure}

Note that centering the unit-information prior on \(\hat\theta\) amounts
to using the data twice (i.e., once to inform the prior distribution
under \(\mathcal{H}_1\) and then again to test it) and will bias the
evidence in favor of \(\mathcal{H}_1\)---particularly in small samples
(see also \citeproc{ref-Bickel_2025}{Bickel, 2025}). The result may be
considered a lower bound on Bayes factors for the unit-information prior
(\citeproc{ref-Held2018}{Held \& Ott, 2018}). When the prior is, more
defensibly, centered on the test value \(\theta_0\) the bias in favor of
\(\mathcal{H}_1\) is

\begin{equation}\protect\phantomsection\label{eq-upi-jab-bias}{
k = \sqrt{1 + 1/n_\text{eff}} \; \exp\left\{-\frac{1}{2}\, \frac{(\hat\theta - \theta_0)^2}{\sigma^2} \; \frac{1}{n_\text{eff} + 1} \right\},
}\end{equation} with \(\sigma\) denoting the asymptotic standard
deviation coefficient, such that the standard error of the maximum
likelihood estimator can be expressed as
\(\text{SE}(\hat\theta) = \sigma / \sqrt{n_\text{eff}}\)
(Appendix~\ref{sec-neff}). Solving for \(n_\text{eff}\) yields the
following multi-valued function that can be used to determine the
minimum effective sample size required to limit the bias to a desired
level,

\begin{equation}\protect\phantomsection\label{eq-upi-jab-bias-min-n}{
n_{\text{eff}} = \frac{-\mathcal{W}\left\{ -\frac{\hat\delta^2}{k^2} \exp\left(-\hat\delta^2\right) \right\}}{\hat\delta^2 + \mathcal{W}\left\{ -\frac{\hat\delta^2}{k^2} \exp\left(-\hat\delta^2\right) \right\}},
}\end{equation} where \(\mathcal{W}()\) is the Lambert \(\mathcal{W}\)
function and the standardized effect size \(\hat\delta\) is defined as
\((\hat\theta - \theta_0) / \sigma\). We provide the proof in
Appendix~\ref{sec-min-n-bias}. Both branches of the Lambert
\(\mathcal{W}\) function can apply but only branches yielding positive
\(n_\text{eff}\) should be used. Figure~\ref{fig-jab-bias-contour} shows
how bias relates to effective sample size and the standardized effect
size. A first-order Taylor expansion at \(1/n_\text{eff} = 0\)
(Appendix~\ref{sec-min-n-bias}) gives a simpler large-sample
approximation that helps to clarify the relationship:

\[
n_{\text{eff}} \approx \frac{1 - \hat\delta^2}{2 \log(k)}.
\] This simple form reveals that JAB is biased only towards
\(\mathcal{H}_1\) (\(k > 1\)) if \(\hat\delta \lessapprox 1\); JAB is
biased towards \(\mathcal{H}_0\) (\(k < 1\)) for larger effects. The
bias towards \(\mathcal{H}_1\) is maximal when the observed effect is
\(\hat\delta = 0\) so the minimum sample size is
\(n_\text{eff} \approx 1/\{2 \log(k)\}\). The corresponding effect size
for the same level of bias towards \(\mathcal{H}_0\) is given by
\(\hat\delta \approx \sqrt{1 - 2 \; n_\text{eff} \; \log(k^*)}\) with
\(k^* = 1/k\). For example, to limit the bias towards \(\mathcal{H}_1\)
to \(k = 1.10\) (i.e.~10\%) requires an minimum sample size of
\(n_\text{eff} \approx 1 / \{2 \, \log(1.10)\} = 5.25\). At this sample
size, the bias towards \(\mathcal{H}_0\) is smaller than \(k = 1/1.10\)
for any
\(\hat\delta \lessapprox \sqrt{1 - (2 \cdot 5.25) \; \log(1 / 1.10)} = 1.41\).

The dashed lines in Figure~\ref{fig-jab-bias-contour} show that this
approximation is quite accurate across a large range of effect sizes.
Only for large effects does the approximation overstate the required
minimum sample size and may be considered a conservative estimate. For
comparison, the exact minimum sample size to limit the bias towards
\(\mathcal{H}_1\) to \(k = 1.10\) (i.e.~10\%) is
\(n_\text{eff} = 4.76\), which also limits the bias towards
\(\mathcal{H}_0\) for effects of \(\hat\delta \leq 1.48\). An effect of
this magnitude provides strong evidence against \(\mathcal{H}_0\),
\(\text{BF}_{01} = 1/31.40\) and \(\text{JAB}_{01} < 1/28.54\). Taken
together these results show that JAB with the unit-information prior
provides an accurate approximation of the corresponding analytic Bayes
factor (Equation~\ref{eq-analytic-bf}) even in very small samples and
for a wide range of effect sizes.

As an aside, these results are also relevant to a recently suggested
small-sample correction of the BIC suggested by Bickel
(\citeproc{ref-Bickel_2025}{2025}) This is because JAB with the
unit-information prior centered on the test value \(\theta_0\) provides
a theoretical justification for the ``evidential'' BIC (EvBIC), which
raises the likelihood to the power of
\(1 - 1/n_\text{eff} = (n_\text{eff}-1)/n_\text{eff}\) to correct the
evidential bias in the Bayes factor approximation based on BIC,
Figure~\ref{fig-p-jab} A. When models differ by one scalar parameter,
the formula for the Bayes factor is

\[
\begin{aligned}
\text{BF}^\text{EvBIC}_{01} & = \sqrt{n_\text{eff}} \; \left( \mathcal{L}_{0} / \mathcal{L}_{1} \right)^{1 - 1/n_\text{eff}} \\
& = \sqrt{n_\text{eff}} \; \exp\left[-0.5 \; (n_\text{eff} - 1)/n_\text{eff} \; \left\{2 \, \log(\mathcal{L}_{1} / \mathcal{L}_{0})\right\} \right] \\
& = \sqrt{n_\text{eff}} \; \exp\left\{-0.5 \; (n_\text{eff} - 1)/n_\text{eff} \; F^{-1}_{\chi^2(1)}(1-p) \right\} \\
& = \text{JAB}_{01},
\end{aligned}
\] exactly with a \(p\)-value from a likelihood ratio test and for the
unit-information prior centered on the test value \(\theta_0\),
Equation~\ref{eq-jab-uip}. This expression shows that the small-sample
correction of raising the likelihood to the power of
\((n_\text{eff}-1)/n_\text{eff}\) is equivalent to centering the
unit-information prior on the \(\theta_0\) rather than the maximum
likelihood estimate \(\hat\theta\)
(\citeproc{ref-Wagenmakers2022}{Wagenmakers, 2022}, Eq. 5),

\[
\text{JAB}_{01}^{(\hat\theta)} = \sqrt{n_\text{eff}} \; \exp\left\{ -0.5 \; F^{-1}_{\chi^2(1)}(1-p) \right\}.
\]

\section{The relationship between surprise and
evidence}\label{sec-evidence-in-p}

Grades of surprise have been suggested to aid the evidential
interpretations of the \(p\)-value (\citeproc{ref-Bland2015}{Bland,
2015, p. 117}; \citeproc{ref-Cox2011}{Cox \& Donnelly, 2011};
\citeproc{ref-Muff_2022}{Muff et al., 2022};
\citeproc{ref-Wasserman2004}{Wasserman, 2004, p. 157}).
Table~\ref{tbl-evidence-categories} lists suggested labels for ranges of
\(p\)-values (also see \citeproc{ref-Held2018}{Held \& Ott, 2018}).
Three aspects are worth noting. First, the grades of surprise correspond
to typical thresholds of \(p\)-values. Second, the grades of surprise
are asymmetric: \(p > .10\) is said to provide ``little or no evidence''
against \(\mathcal{H}_0\) and only \(p < .10\) is considered noteworthy;
no \(p\)-value is considered to provide evidence \emph{in favor} of
\(\mathcal{H}_0\). Third, the grades of surprise are independent of
sample size.

\begin{landscape}

\begin{table}

{\caption{{Categorical interpretations of two-sided \(p\)-values as
surprise (evidence against \(\mathcal{H}_0\)) and corresponding
approximate Bayes factors
(evidence).\newline}{\label{tbl-evidence-categories}}}
\vspace{-20pt}}

\centering
\begin{threeparttable}
\begin{tabular}[t]{lllcccc}
\toprule
\multicolumn{1}{c}{ } & \multicolumn{2}{c}{Grades of surprise} & \multicolumn{1}{c}{ } & \multicolumn{3}{c}{Bayesian evidence} \\
\cmidrule(l{3pt}r{3pt}){2-3} \cmidrule(l{3pt}r{3pt}){5-7}
$p$ & Bland (2015) & Wasserman (2004) & $s$ & $\text{JAB}_{10}(n_\text{eff} = 8)$ & $\max(\text{BF}^{\mathcal{N}}_{10})$ & $\max(\mathcal{L}_1 / \mathcal{L}_0)$\\
\midrule
$(1.000, .100]$ & Little or no surprise & Little or no surprise & $(0.00, 3.32]$ & $(0.35, 1.15]$ & $(1.00, 1.43]$ & $(1.00, 3.87]$\\
$(.100, .050]$ & Weak surprise & Weak surprise & $(3.32, 4.32]$ & $(1.15, 1.90]$ & $(1.43, 2.11]$ & $(3.87, 6.83]$\\
$(.050, .010]$ & Surprise & Strong surprise & $(4.32, 6.64]$ & $(1.90, 6.44]$ & $(2.11, 6.50]$ & $(6.83, 27.59]$\\
$(.010, .001]$ & Strong surprise & Very strong surprise & $(6.64, 9.97]$ & $(6.44, 40.34]$ & $(6.50, 41.38]$ & $(27.59, 224.48]$\\
$(.001, .000)$ & Very strong surprise &  & $(9.97, \infty)$ & $(40.34, \infty)$ & $(41.38, \infty)$ & $(224.48, \infty)$\\
\bottomrule
\end{tabular}
\begin{tablenotes}
\small
\item [] $\text{JAB}_{10}$ assuming a unit-information prior centered on the test-value $\theta_0$. $s = -\log_2(p)$ is the surprise in bits (\citeproc{ref-Rafi2020}{Rafi \& Greenland, 2020}). $\max(\text{BF}^{\mathcal{N}}_{10})$ is the upper limit on the evidence for $\mathcal{H}_1$ when the prior distribution is normal (\citeproc{ref-Edwards1963}{Edwards et al., 1963, p. 231}).
\end{tablenotes}
\end{threeparttable}

\end{table}

\end{landscape}

\begin{figure}

\caption{\label{fig-p-jab}Relationship between two-sided \(p\)-value
(surprise) and \(\text{JAB}_{10}\) (evidence) as a function of effective
sample size \(n_\text{eff}\). Dark, solid, horizontal lines indicate
suggested grades of surprise (e.g., \citeproc{ref-Bland2015}{Bland,
2015}; \citeproc{ref-Wasserman2004}{Wasserman, 2004}).
\(\text{JAB}_{10}\) is given for the unit-information prior centered on
the test-value \(\theta_0\). The color gradient shows the evidence and
solid contour lines indicating suggested grades of evidence
(\citeproc{ref-Lee2014}{Lee \& Wagenmakers, 2014};
\citeproc{ref-Wasserman2000}{Wasserman, 2000}). \textbf{A} Relationship
across a large range of \(n_\text{eff}\) on logarithmic scale. The
dotted lines represent the \(\text{JAB}_{10}\) for the unit-information
prior centered on the maximum likelihood estimate \(\hat\theta\).
\textbf{B} Shift of the grades of evidence implied by a \(p\)-value from
typically small (\(n < 50\)) to large effective sample sizes
(\(n > 250\)).}

\centering{

\pandocbounded{\includegraphics[keepaspectratio]{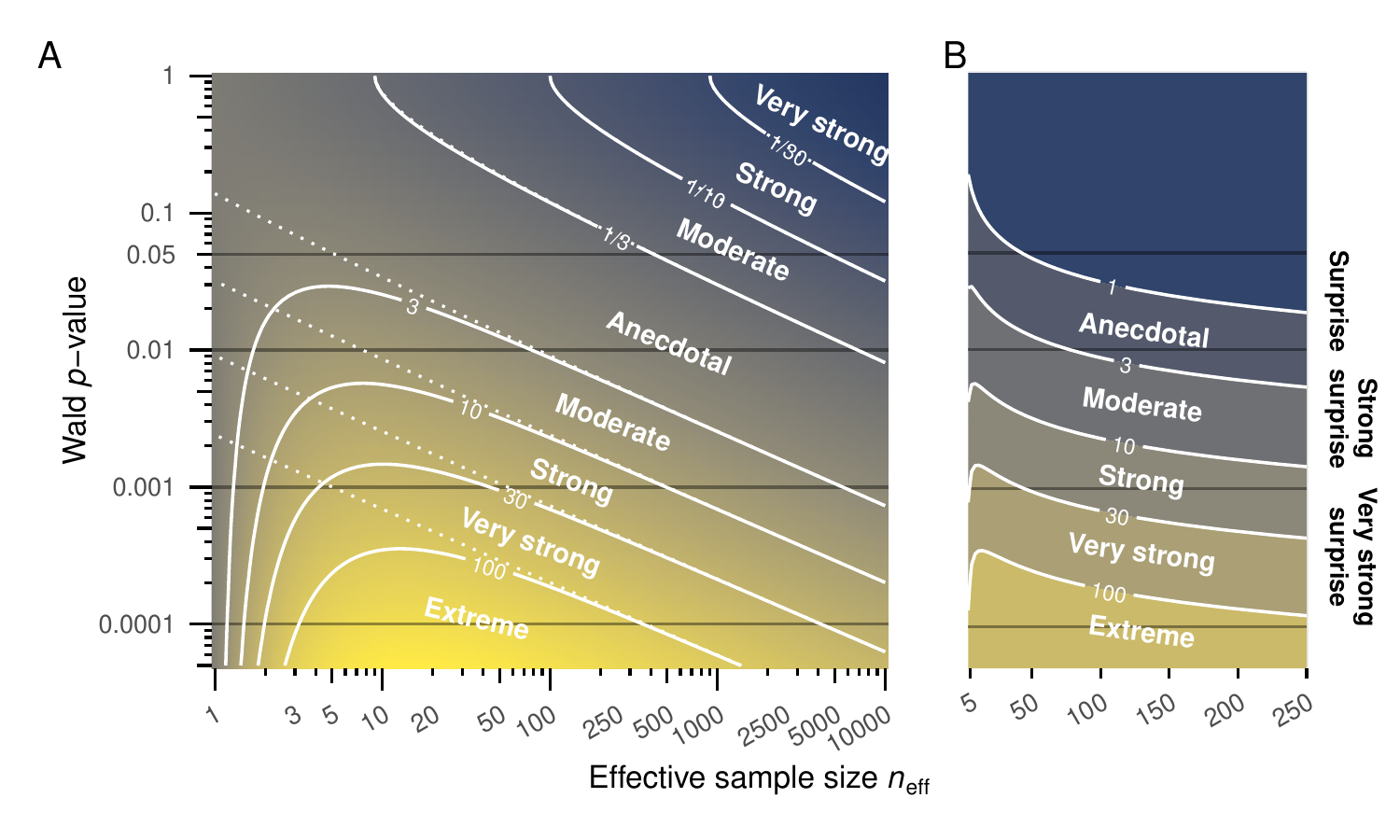}}

}

\end{figure}

JAB can be used relate these grades of surprise to Bayesian evidence
from objective hypothesis tests. Figure~\ref{fig-p-jab} A shows how the
\(p\)-value relates to \(\text{JAB}_{10}\) as a function of the
effective sample size. Solid white lines represent suggested grades of
evidence (\citeproc{ref-Jeffreys1961}{Jeffreys, 1961};
\citeproc{ref-Lee2014}{Lee \& Wagenmakers, 2014};
\citeproc{ref-Wasserman2000}{Wasserman, 2000}). Note that both axes are
on the log scale. We relate surprise to \(\text{JAB}_{10}\) for a common
prior choice in objective testing: the unit-information prior centered
on the test value \(\theta_0\)
(\citeproc{ref-Wagenmakers2022}{Wagenmakers, 2022}, Ep. 4), that is,

\begin{equation}\protect\phantomsection\label{eq-jab-uip}{
\text{JAB}_{01} = \sqrt{n_\text{eff}} \; \exp\left\{ -0.5 \; (n_\text{eff}-1) / n_\text{eff} \; F^{-1}_{\chi^2(1)}(1-p) \right\}.
}\end{equation} As we showed in the previous section, this prior is
sufficiently wide to yield an acceptable approximations of the analytic
Bayes factor even in small samples.

Several observations can be made about the relationship between the
surprise expressed in the \(p\)-value and the evidence as quantified by
the Bayes factor. First, it is clear from Figure~\ref{fig-p-jab} A that
the Bayes factor implied by a fixed \(p\)-value varies with the
effective sample size. Fixed \(p\)-values are represented by dark
horizontal lines. In contrast, in small samples the \(\text{JAB}_{10}\)
increases with the effective sample size
(\(n_\text{eff} \lessapprox 10\)) but then decreases, as shown in
Figure~\ref{fig-p-jab} B and highlighted in Figure~\ref{fig-p-jab} A.
Consequently, \(p = .01\) may constitute moderate evidence against
\(\mathcal{H}_0\) when \(n_\text{eff} = 50\)
(\(\text{JAB}_{10} = 3.65\)), but moderate evidence \emph{in favor} of
\(\mathcal{H}_0\) when \(n_\text{eff} = 10{,}000\),
\(\text{JAB}_{10} = 1/3.63\). This is an example of the Jeffreys-Lindley
paradox (\citeproc{ref-Wagenmakers_2022}{Wagenmakers \& Ly, 2022}). The
fact that the evidence implied by a \(p\)-value depends on sample size
implies that there is no simple relationship between the grades of
surprise and analogous proposals of grades of evidence
(\citeproc{ref-Lee2014}{Lee \& Wagenmakers, 2014};
\citeproc{ref-Wasserman2000}{Wasserman, 2000}). We find good agreement
between grades of surprise and grades of evidence for an effective
sample size of \(n_\text{eff} \approx 8\): For example, strong surprise
(\(p = .005\)) coincides with strong evidence,
\(\text{JAB}_{10} = 11.11\). But note that this is a cherry-picked point
of comparison: At \(n_\text{eff} \approx 8\) the Bayes factor is near
the theoretical upper bound \(\max(\text{BF}^{\mathcal{N}}_{10})\) for
this class of prior distribution, Table~\ref{tbl-evidence-categories},
(\citeproc{ref-Edwards1963}{Edwards et al., 1963, p. 231}). As effective
sample size decreases or increases, strong surprise (\(p = .005\))
corresponds to decreasing evidence. For example, moving from
\(n_\text{eff} = 8\) to \(n_\text{eff} = 80\) halves the evidence,
\(\text{JAB}_{10} = 5.47\), turning strong into moderate evidence---a
shift by a full grade. The solid white lines in Figure~\ref{fig-p-jab}
show that for \(n_\text{eff} > 20\) the surprise \(-\log(p)\) has to
increase approximately linearly for the implied evidence to remain
constant (\citeproc{ref-Benjamin_2017}{Benjamin et al., 2017}). This
means that changes to the effective sample size strongly affect the
evidence in small samples, but become less consequential as the
effective sample size increases. This decelerating decrease in the
evidence implied by a fixed \(p\)-value can be seen more clearly when
effective sample size is shown in its natural scale,
Figure~\ref{fig-p-jab} B. The solid white lines show that for fixed
evidence the surprise \(-\log(p)\) decreases more rapidly when effective
sample size increases from 5 to 20 than when it increases from 235 to
250. This shows that the relationship between surprise and evidence
depends on effective sample size and that surprise is generally larger
than the implied evidence for \(\mathcal{H}_1\) relative to
\(\mathcal{H}_0\).

The second observation pertains to the effect that the location of the
unit-information prior has on the relationship between surprise and
evidence in small samples. Centering the unit-information prior on the
maximum likelihood estimate \(\hat\theta\) yields
(\citeproc{ref-Wagenmakers2022}{Wagenmakers, 2022}, Eq.~5)

\begin{equation}\protect\phantomsection\label{eq-jab-uip-mle}{
\begin{aligned}
\text{JAB}_{01}^{(\hat\theta)} & = \sqrt{n_\text{eff}} \; \exp\left\{ -0.5 \; F^{-1}_{\chi^2(1)}(1-p) \right\}, \\
\end{aligned}
}\end{equation} which is equal to the Bayes factor approximation based
on \(\Delta\text{BIC}\) (\citeproc{ref-Wagenmakers_2007}{Wagenmakers,
2007}),

\begin{equation}\protect\phantomsection\label{eq-jab-bic}{
\begin{aligned}
\text{BF}_{01} & \approx \exp\left(-0.5 \; \Delta\text{BIC}_{01}\right) \\
& = \exp\left[-0.5 \; \left\{\Lambda - \log(n_\text{eff})\right\}\right] \\
& = \sqrt{n_\text{eff}} \; \exp\left\{ -0.5 \; F^{-1}_{\chi^2(1)}(1-p) \right\} \\
& = \text{JAB}_{01}^{(\hat\theta)}
\end{aligned}
}\end{equation} exactly with a \(p\)-value from a likelihood ratio test
and \(\Lambda = 2 \; \log(\mathcal{L}_{1} / \mathcal{L}_{0})\). This
pseudoprior amounts to using the data twice---to inform the prior
distribution under \(\mathcal{H}_1\) and to test \(\mathcal{H}_1\). The
dotted lines in Figure~\ref{fig-p-jab} A show the relationship between
this \(\text{JAB}_{10}^{(\hat\theta)}\) and \(p\) as a function of
\(n_\text{eff}\). Comparing the dotted to the solid lines shows that, in
small samples, \(\text{JAB}_{10}^{(\hat\theta)}\) overstates the
evidence against \(\mathcal{H}_0\). As the effective sample size
decreases, the dotted lines continue to climb approximately linearly,
whereas the solid lines slope downwards. However, this divergence
decreases as \(p\) or the effective sample size increases,

\[
\frac{\text{JAB}_{01}^{(\hat\theta)}}{\text{JAB}_{01}} = \exp\left\{ -0.5 \; F^{-1}_{\chi^2(1)}(1-p) \right\}^{1/n_\text{eff}}.
\] For example, at modest \(n_\text{eff} = 10\) and strong surprise of
\(p = .005\), \(\text{JAB}_{10}^{(\hat\theta)}\) overstates the evidence
against \(\mathcal{H}_0\) by a factor of
\(\text{JAB}_{01}^{(\hat\theta)} / \text{JAB}_{01} = 16.26 / 10.96= 1.48\).

A third observation from Figure~\ref{fig-p-jab} A pertains to the
evidence corresponding to \(p > .1\). In line with the interpretation of
\(p\) as surprise, in very small samples (\(n_\text{eff} < 9\)) such
results correspond to little or no evidence either against
\(\mathcal{H}_0\) or for \(\mathcal{H}_1\). In moderate
(\(9 < n_\text{eff} < 100\)) and large samples
(\(100 < n_\text{eff} < 900\)), however, \(p > .1\) may indeed
correspond to moderate or even strong evidence \emph{in favor of}
\(\mathcal{H}_0\). From a Bayesian perspective, the lack of surprise
implied by \(.1 < p < 1\) can correspond to meaningful evidence in favor
of \(\mathcal{H}_0\), if the sample is sufficiently large.

Fourth, JAB can be used to derive lower and upper bounds on the Bayes
factor. For simplicity, consider \(\text{JAB}_{01}^{(\hat\theta)}\). The
lower bound of the evidence for a given \(p\) is reached when
\(n_\text{eff} = 1\),

\[
\min(\text{JAB}_{01}^{(\hat\theta)}) = \exp\left\{-0.5 \; F^{-1}_{\chi^2(1)}(1-p)\right\}.
\] This is the lower bound on the likelihood ratio,
Table~\ref{tbl-evidence-categories}, (\citeproc{ref-Berger1987}{Berger
\& Sellke, 1987, p. 116}; \citeproc{ref-Edwards1963}{Edwards et al.,
1963, p. 228}). For a comprehensive review on other methods to obtain
lower bounds on the Bayes factor see Held and Ott
(\citeproc{ref-Held2018}{2018}). Conversely, the upper bound on the
evidence for \(\mathcal{H}_0\) for a given effective sample size
\(n_\text{eff}\) is reached when \(p \to 1\) and thus

\begin{equation}\protect\phantomsection\label{eq-max-jab}{
\max(\text{JAB}_{01}) = \sqrt{n_\text{eff}}.
}\end{equation}

This simple expression holds regardless of the center of the
unit-information prior and is a useful reference point when planning
studies or evaluating claims about the absence of an effect based on a
large \(p\)-value.

It is clear from Equation~\ref{eq-jab} that the relationship between the
\(p\)-value and the Bayes factor depends on the prior distribution under
\(\mathcal{H}_1\). The results for the unit-information prior are
attractive because of its popularity in objective Bayesian hypothesis
testing, but more narrow priors are appropriate in the presence of
strong background knowledge. The effect of different prior distributions
on \(\text{JAB}_{01}\) can be explored through a prior sensitivity
analysis (\citeproc{ref-Sinharay_2002}{Sinharay \& Stern, 2002}).
Changing the prior simply scales \(\text{JAB}_{0i}\) by the inverse
ratio of the prior densities at \(\hat\theta\)
(\citeproc{ref-Ly_2022}{Ly \& Wagenmakers, 2022}):

\[
\begin{aligned}
\frac{\text{JAB}_{02}}{\text{JAB}_{01}} & = \frac{A_2}{A_1} = \frac{g(\hat \theta \mid \mathcal{H}_1)}{g(\hat \theta \mid \mathcal{H}_2)} \\
\text{JAB}_{02} & = \text{JAB}_{01} \; \frac{g(\hat \theta \mid \mathcal{H}_1)}{g(\hat \theta \mid \mathcal{H}_2)}.
\end{aligned}
\] Therefore, the ratio of the prior densities at \(\hat\theta\)
quantifies the change in evidence caused by assuming a different prior
distribution. This ratio may also be thought of as a measure of relative
prior-data conflict (\citeproc{ref-Young_1996}{Young \& Pettit, 1996}).
It is worth repeating here that JAB is most accurate for objective
uninformed prior distributions unless the sample size is large. Hence,
this approximate prior sensitivity (or prior-data conflict) analysis is
best suited to compare relatively wide prior distributions that place
non-negligible mass near \(\hat\theta\) in large samples.

\section{Concluding Comments}\label{concluding-comments}

By relating the \(p\)-value to Jeffreys's approximate Bayes factor (JAB)
we have bridged the gap between two seemingly incompatible statistical
concepts: the surprise quantified by the \(p\)-value and the relative
evidence quantified by the Bayes factor (BF) from an objective Bayesian
hypothesis test. Our investigation of JAB, with the unit-information
prior as a running example, shows that

\begin{enumerate}
\def\labelenumi{\arabic{enumi}.}
\tightlist
\item
  JAB approximates the BF well when the sampling distribution of the
  maximum likelihood estimate is approximately normal and it is used
  with objective, relatively vague priors or in large samples.
\item
  even in small samples (e.g., \(n_\text{eff} = 5\)), the
  unit-information prior yields an accurate approximation. A reanalysis
  of 704 published \(t\)-tests showed that, in practice, JAB also gives
  a fair approximation to the commonly used default JZS-Bayes factors
  (Figure~\ref{fig-jabp-jzs}).
\item
  the relative evidence implied by a \(p\)-value depends on sample size.
  For example, with a unit-information prior a \(p = .05\) at
  \(n_\text{eff} = 100\) yields \(\text{JAB}_{01} = 1.5\), meaning weak
  support in favor of the null hypothesis.
\item
  common verbal scales (e.g., \(p < . 05\) = ``strong surprise,''
  \citeproc{ref-Wasserman2004}{Wasserman, 2004}) do not translate to
  stable Bayesian evidence categories. For the unit-information prior,
  they only align with at \(n_\text{eff} \approx 8\), at which point the
  Bayes factor approaches its theoretical upper bound
  (Table~\ref{tbl-evidence-categories}). At common sample sizes in the
  empirical sciences of \(2 < n_\text{eff} < 200\),
  \(p \in [.05, .001]\) often fall into lower Bayesian evidence
  categories (e.g., ``Anecdotal evidence'' rather than ``Strong
  surprise'', Figure~\ref{fig-jabp-jzs}).
\item
  for the unit-information prior and \(n_\text{eff} > 100\), \(p > .1\)
  corresponds to moderate or strong evidence for \(\mathcal{H}_0\)
  relative to \(\mathcal{H}_1\)---contradicting the common heuristic
  that non-significance is evidentially uninformative
  (Figure~\ref{fig-jabp-jzs}). The upper bound
  \(\max(\text{JAB}_{01}) = \sqrt{n_\text{eff}}\) provides a reference
  point (Equation~\ref{eq-max-jab}).
\end{enumerate}

Finally, we believe that JAB also has practical utility. JAB can help to
distinguish between a large \(p\)-value that indicates the absence of an
effect and one reflecting ambiguous results. A Bayes factor close to 1,
indicating that the data are equally likely under \(\mathcal{H}_0\) and
\(\mathcal{H}_1\), should prompt researchers to avoid strong claims and
collect additional data or design a more informative study. Hence, JAB
can be a useful supplement to \(p\)-values, one that can be conveniently
calculated from routinely reported statistics.

Readers of the scientific literature may particularly benefit from
expressions that allow quick back-of-the-envelope calculations to
evaluate a reported result. The upper bound
\(\max(\text{JAB}_{01}) = \sqrt{n_\text{eff}}\) may be used to evaluate
claims about the absence of effects. Wagenmakers
(\citeproc{ref-Wagenmakers2022}{2022}, Eq. 9) used JAB to develop the
\(3p\sqrt{n}\)-rule, an even simpler approximation that requires only
\(p\) and \(n_\text{eff}\) (also see
\citeproc{ref-KagerMeesterinpress}{Kager \& Meester, 2026}),

\[
\text{JAB}_{01}^{(\hat\theta)} \approx
\begin{cases}
\begin{aligned}
3 p & \sqrt{n_\text{eff}} & \quad \text{if} \; \phantom{.10 < } & p \leq .10 \\
\sqrt{p} & \sqrt{n_\text{eff}} & \text{if} \; .10 < & p \leq .50 \quad 
\text{(simpler)} \\
\; \frac{4}{3} p^{2/3} & \sqrt{n_\text{eff}} & \text{if}  \; .10 < & p \leq .50 \quad \text{(more precise)} \\
p^{1/4} & \sqrt{n_\text{eff}} & \text{if} \; \phantom{.10 < } & p > .50,
\end{aligned}
\end{cases}
\] for a unit-information prior centered on the maximum likelihood
estimate \(\hat\theta\). And because it approximates Bayes factors, the
evidence expressed in JAB provides a valid estimate of the evidence even
when the data were sampled until \(p < \alpha\) (optional stopping). So,
JAB can be a computationally cheap alternative to a full sequential
Bayesian analysis, i.e., when continuously monitoring the evidence as
the data come in (\citeproc{ref-de_Heide_2020}{Heide \& Grünwald, 2020};
\citeproc{ref-Ramdas2023}{Ramdas et al., 2023};
\citeproc{ref-Rouder_2014}{Rouder, 2014};
\citeproc{ref-Wagenmakers2026}{Wagenmakers et al., 2026}). Taken
together this is why we believe that JAB can help authors, reviewers,
and readers of the scientific literature better interpret empirical
results.

\section*{Disclosure statement}\label{disclosure-statement}
\addcontentsline{toc}{section}{Disclosure statement}

No external funding was received for this work. The authors declare that
they have no financial or competing interests.

\section{Declaration of generative AI
use}\label{declaration-of-generative-ai-use}

OpenAI's ChatGPT 5.5 and Google's Gemini Pro 3.1 were used in part to
derive and verify proofs. The initial draft of this manuscript was
written without the use of Generative AI. Anthropic's Claude 5 Sonnet,
OpenAI's ChatGPT 5.5, and Google's Gemini Pro 3.1 were used to review
the manuscript for clarity, conciseness, grammar, accuracy, and
consistency. All content was verfied by the authors.

\clearpage

\section{References}\label{references}

\protect\phantomsection\label{refs}

\appendix

\renewcommand{\thesubsection}{\Alph{appendix}.\arabic{subsection}}

\section{\texorpdfstring{What is Effective Sample Size
\(n_\text{eff}\)?}{What is Effective Sample Size n\_\textbackslash text\{eff\}?}}\label{sec-neff}

In the main text, we largely glossed over an important issue: in
Equation~\ref{eq-jab}, what is the effective sample size
\(n_\text{eff}\)? The term \(\sqrt{n_\text{eff}}\) is the denominator
after factoring the standard error into
\(\text{SE}(\hat\theta) = \sigma / \sqrt{n_\text{eff}}\), where
\(\sigma\) is independent of sample size. \(n_\text{eff}\) is a function
of sample size and scales the standard deviation of the asymptotic
sampling distribution of \(\hat \theta\). Hence, the correct definition
of \(n_\text{eff}\) depends on \(\theta\) and is calculated differently
for each model and parameterization. Determining the correct
\(n_\text{eff}\) can be difficult in more complex models. In this case
it may be more convenient to use Equation~\ref{eq-jab} with the standard
error \(\text{SE}(\hat \theta)\) included:

\[
\text{JAB}_{01} = [\sqrt{2 \pi}~\text{SE}(\hat\theta)~g(\hat \theta \given \mathcal{H}_1)]^{-1} \; \exp(-0.5 \; W).
\]

For the purpose of this paper, factoring the the standard error
clarifies that the relationship between \(p\)-value and Bayes factor
depends on sample size. In the following, we show a simple derivation of
\(n_\text{eff}\) for the applications shown in Figure~\ref{fig-jabp-jzs}
and Figure~\ref{fig-jab-prop}. Berger et al.
(\citeproc{ref-Berger2013}{2013}) provide a general treatment of
effective sample size in the linear model (also see
\citeproc{ref-Pauler1998}{Pauler, 1998}).

\subsection{\texorpdfstring{One-sample \(t\)-test of the
mean}{One-sample t-test of the mean}}\label{one-sample-t-test-of-the-mean}

In the case of a one-sample \(t\)-test, \(\hat \theta = \hat \mu\)---the
sample mean---and the standard error is
\(\text{SE}(\hat \theta = \hat \mu) = \hat \sigma / \sqrt{n}\). Here,
the effective sample size is simply the number of observations,
\(n_\text{eff} = n\), and \(\hat\sigma\) is the estimated population
standard deviation. Similarly, in the paired-sample \(t\)-test,
\(\hat \theta = \hat \mu_\Delta\)---the sample mean of the differences
between the paired observations---and the standard error is
\(\text{SE}(\hat \theta = \hat \mu_\Delta) = \hat \sigma_\Delta / \sqrt{n_\Delta}\).
Now, the effective sample size is the number of differences or,
equivalently, the number of pairs, \(n_\text{eff} = n_\Delta\).

\subsection{\texorpdfstring{Two-sample \(t\)-tests of
means}{Two-sample t-tests of means}}\label{two-sample-t-tests-of-means}

In the independent sample \(t\)-test,
\(\hat \theta = \hat \mu_1 - \hat \mu_2 = \Delta \hat \mu\)---the
difference between the sample means---and the standard error is

\[
\begin{aligned}
\text{SE}(\hat \theta = \Delta \hat \mu) & = \hat \sigma_p \; \sqrt{1/n_1 + 1/n_2} \\
& = \frac{ \hat \sigma_p }{ \sqrt{(n_1 \; n_2) / (n_1 + n_2)} },
\end{aligned}
\] where \(\hat \sigma_p\) is the pooled estimate of the population
standard deviation, i.e.~the assumedly \emph{common} standard deviation
estimated using the data from both samples of size \(n_1\) and \(n_2\).
So here, the effective sample size \(n_\text{eff}\) is half the harmonic
mean of the two sample sizes,

\[
n_\text{eff} = H(n_1, n_2) / 2 = (n_1 \; n_2) / (n_1 + n_2),
\] an average dominated by the smaller sample size. In balanced designs,
where \(n = n_1 = n_2\), this expression simplifies to
\(n_\text{eff} = n / 2\), i.e., half the number of observations per
group.

\subsection{\texorpdfstring{Two-sample \(z\)-test of
proportions}{Two-sample z-test of proportions}}\label{two-sample-z-test-of-proportions}

For the two-proportion \(z\)-test (and Pearson's \(\chi^2\)-test for
\(2\times\)-contigency tables), the effective sample size can be derived
from the formula for the asymptotic standard error of the assumedly
common proportion \(\widehat{p_p} = (y_1 + y_2) / (n_1 + n_2)\)
(\citeproc{ref-Agresti_2006}{Agresti, 2006, p. 26}),

\[
\text{SE}(\hat\theta = \widehat{p_p}) = \sqrt{ \widehat{p_p} (1 - \widehat{p_p}) } \, \sqrt{ 1/n_1 + 1/n_2 }.
\] Hence, as for the comparison of independent means,
\(n_\text{eff} = (n_1 \; n_2) / (n_1 + n_2)\). Similarly, for the test
of log odds ratio the large-sample approximation of the standard error
(\citeproc{ref-Agresti_2006}{Agresti, 2006, p. 30}),

\[
\begin{aligned}
\text{SE}(\hat{\theta} = \log{\widehat{\text{OR}_p}}) & = \sqrt{ \frac{1}{y_1} + \frac{1}{y_2} + \frac{1}{n_1 - y_1} + \frac{1}{n_2 - y_2} } \\
& = \sqrt{ \frac{1}{n_1 \, \widehat{p_p}} + \frac{1}{n_2 \, \widehat{p_p}} + \frac{1}{n_1 (1 - \widehat{p_p})} + \frac{1}{n_2 (1 - \widehat{p_p})} } \\
& =  \sqrt{ \frac{1}{\widehat{p_p} (1 - \widehat{p_p})} } \, \sqrt{ 1/n_1 + 1/n_2 }.
\end{aligned}
\]

\section{\texorpdfstring{Likelihood ratio approximation for one- and
two-sample \(t\)-tests of
means}{Likelihood ratio approximation for one- and two-sample t-tests of means}}\label{sec-llr}

\setcounter{subsection}{0}

In Equation~\ref{eq-jab}, \(W\) is used to approximate the scaled log
likelihood ratio
\(\Lambda = 2 \; \log(\mathcal{L}_{1} / \mathcal{L}_{0})\), because
\(W\) is asymptotically equivalent to \(\Lambda\) under standard
regularity conditions (\citeproc{ref-Buse_1982}{Buse, 1982};
\citeproc{ref-Engle1984}{Engle, 1984, p. 798};
\citeproc{ref-Vaart_1998}{Vaart, 1998, p. 227}). Further,
Equation~\ref{eq-w-p} shows that the Wald statistic \(W\) can be
calculated from its corresponding \(p\)-value. However, the \(p\)-values
shown in Figure~\ref{fig-jabp-jzs} are based on the \(t\)-statistic.
When these \(p\)-values are transformed to an approximate Wald statistic
\(W_t\) to calculate JAB, we deviate from the original derivation of JAB
in two ways: (1) The underlying \(t\)-statistic is based on a standard
error that relies on the unbiased estimate of the population variance
(Bessel's correction, \(n - 1\)). The Wald statistic, however, is based
on the uncorrected maximum likelihood estimate of the variance. (2) The
\(p\)-value is based on the \(t\)-distribution rather than the standard
normal distribution. Figure~\ref{fig-jabp-jzs} illustrates the
consequences of these deviations.

\begin{figure}

\caption{\label{fig-jab-jzs}Linear relationships between JZS-Bayes
factor and \(p\)-value-based JAB for 704 \(t\)-test results collected by
Aczel et al. (\citeproc{ref-Aczel_2018}{2018}) and Wetzels et al.
(\citeproc{ref-Wetzels_2011}{2011}). Triangles represent one-sided
\(p\)-values, circles represent \(\text{JAB}_{01}\), each on logarithmic
scale. The color of points indicates the effective sample size. The
solid grey line shows the estimated linear relationship between
\(p\)-values and Bayes factors.}

\centering{

\pandocbounded{\includegraphics[keepaspectratio]{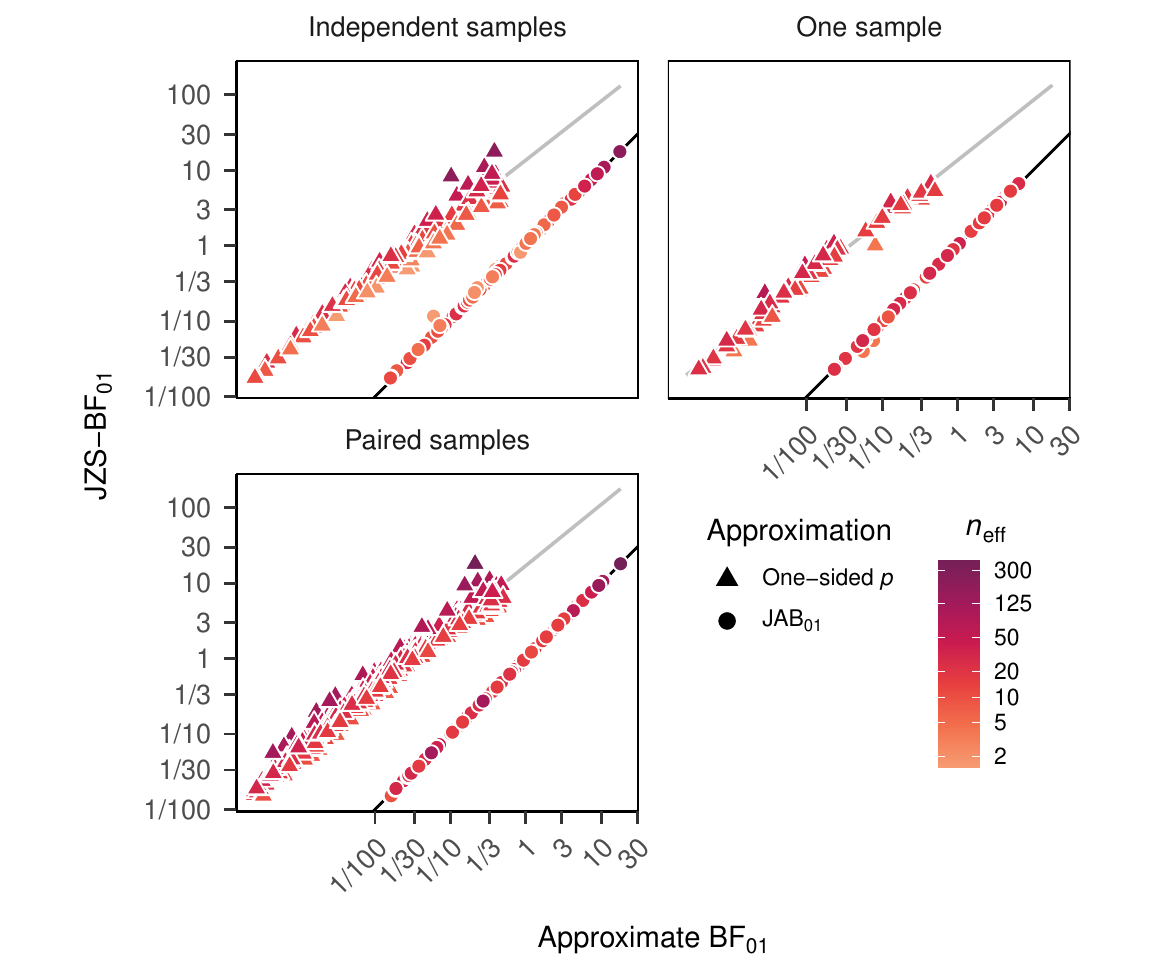}}

}

\end{figure}%

\begin{figure}

\caption{\label{fig-bf-t-z}Bias of likelihood ratio approximations from
\(p\) of a one-sample \(t\)-test (\(W_t\); left panel) and with
additional correction (\(N / (N - 1)\); right panel). The grey area
shows the margin of error of a factor of 3
(\citeproc{ref-Jeffreys1961}{Jeffreys, 1961, p. 433}).}

\centering{

\pandocbounded{\includegraphics[keepaspectratio]{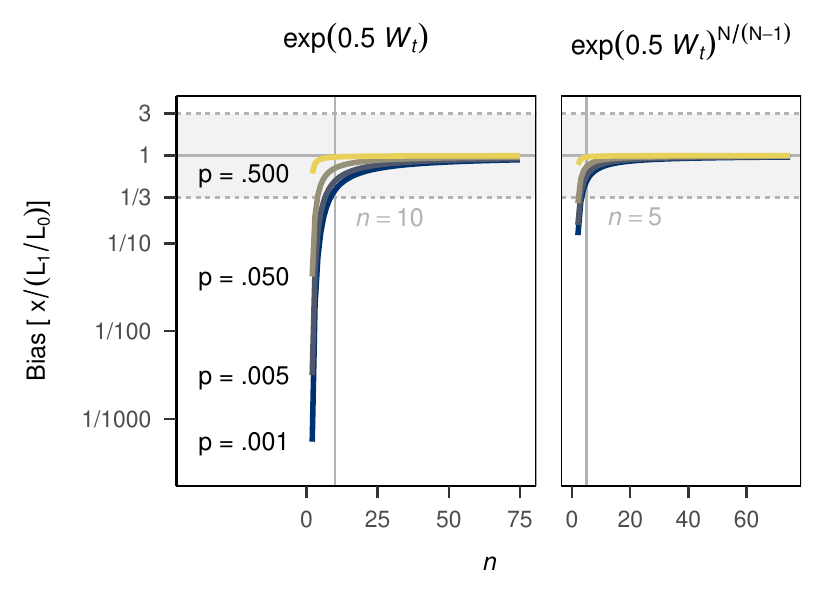}}

}

\end{figure}

To illustrate the quality of the approximation when its assumptions are
met, Figure~\ref{fig-jab-jzs} shows the results using the exact
likelihood ratio, which can be calculated from the \(t\)-value
(\citeproc{ref-Francis2016}{Francis, 2016};
\citeproc{ref-Francis2022}{Francis \& Jakicic, 2022};
\citeproc{ref-Kendall1961}{Kendall \& Stuart, 1961, p. 225};
\citeproc{ref-Murtaugh2014}{Murtaugh, 2014}),

\[
\log(\mathcal{L}_{1} / \mathcal{L}_{0}) = \frac{N}{2}\log\bigg(1 + \frac{t^2}{N - k}\bigg),
\] where \(N\) is the total sample size and \(k\) is the number of
samples. The term \(N - k\) represents the residual degrees of freedom,
which serve as Bessel's correction for the unbiased estimate of the
population variance. As noted above, this is necessary because the
likelihood ratio is based on the uncorrected maximum likelihood estimate
of the variance. We have found that the likelihood ratio approximation
based on \(p\) can similarly be improved if the approximate \(W_t\) is
adjusted by a corrective factor of \(N / (N - k)\) (see also
\citeproc{ref-Bickel_2025}{Bickel, 2025}),

\[
\begin{aligned}
W_t & = \left\{ F^{-1}_{\mathcal{N}(0,1)}(p_t/2) \right\}^2 \\
   & \approx \frac{(\bar y - \mu_0)^2}{\sum{(y_i - \mu_0)^2} / \{(N - k) \; n_\text{eff} \} } \\ \\
W & \approx W_t \frac{N}{N - k} \\
  & \approx \frac{(\bar y - \mu_0)^2}{\sum{(y_i - \mu_0)^2} / (N \; n_\text{eff})},
\end{aligned}
\] where \(y_i\) are samples from the population and \(\mu_0\) is the
population mean under \(\mathcal{H}_0\). Hence,

\[
\text{JAB}_{01} \approx A \; \sqrt{n_\text{eff}} \; \exp\left(-0.5 \; W_t\right)^{N/(N - k)}.
\] The correction yields an approximate \(t\) statistic but calculated
using the maximum likelihood estimate of the variance.
Figure~\ref{fig-bf-t-z} shows the bias in JAB for a one-sample
\(t\)-test when \(W\) approximated from \(p\), with and without the
correction, relative to the analytic likelihood ratio. Both
approximations work well even in relatively small samples---\(n > 10\)
and \(n > 5\), respectively---and in particular for \(p > .05\).

\section{Derivation of JAB}\label{sec-jab-derivation}

\setcounter{subsection}{0}

Jeffreys (\citeproc{ref-Jeffreys1961}{1961}) did not provide a formal
justification for the approximation

\begin{equation}\protect\phantomsection\label{eq-ml-approx}{
\text{p}(\hat\theta \given \mathcal{H}_1) = \int \mathcal{N}(\hat\theta \given \theta, \text{SE}(\hat\theta)^2) \; g(\theta \given \mathcal{H}_1) \; d\theta \approx g(\hat\theta \given \mathcal{H}_1),
}\end{equation} which is central to JAB. Here we try to provide such a
justification with sufficient detail to be understood by a wider
audience.

\subsection{Dirac limit}\label{dirac-limit}

The Dirac limit is one way to justify the approximation. The normal
distribution in Equation~\ref{eq-ml-approx} becomes infinitely
concentrated at the point \(\hat\theta = \theta\) and converges to a
Dirac delta function \(\delta(\theta - \hat\theta)\)---a spike at
\(\hat\theta\)---as \(n \to \infty\) and
\(\text{SE}(\hat\theta) \to 0\). Because of the ``sifting property'' of
the Dirac delta function (\citeproc{ref-Bracewell1999}{Bracewell, 1999,
p. 79}; \citeproc{ref-Roden2014}{Roden, 2014, p. 40}),

\[
\lim_{\text{SE}(\hat\theta) \to 0} \text{p}(\hat\theta \given \mathcal{H}_1) = \int \delta(\theta - \hat\theta) \, g(\theta \given \mathcal{H}_1) \; d\theta = g(\hat\theta \given \mathcal{H}_1)
\] for any prior distribution \(g()\) continuous at \(\hat\theta\). This
shows that the asymptotic approximation applies similarly to other
likelihoods that are asymptotically concentrated. For finite sample
sizes, the quality of the approximation depends on the rate of
concentration.

\subsection{Chib's identity}\label{chibs-identity}

The Dirac limit provides a justification for Equation~\ref{eq-ml-approx}
in the limit of infinite data.

Chib's (\citeproc{ref-Besag_1989}{Besag, 1989};
\citeproc{ref-Chib_1995}{1995}) identity provides an alternative
justification (c.f. \citeproc{ref-Rougier_2020}{Rougier \& Priebe,
2020}). Assuming that
\(\text{p}(\hat\theta \given y, \mathcal{H}_1) > 0\),

\[
\text{p}(y \given \mathcal{H}_1) = \frac{\text{p}(y \given \hat\theta, \mathcal{H}_1)}{\text{p}(\hat\theta \given y, \mathcal{H}_1)} \; g(\hat\theta \given \mathcal{H}_1).
\] The Bayes factor is

\[
\begin{aligned}
\text{BF}_{01} & = \frac{\text{p}(y \given \mathcal{H}_0)}{\text{p}(y \given \mathcal{H}_1)} \\
& = \underbrace{\frac{\text{p}(y \given \theta_0, \mathcal{H}_0)}{\text{p}(y \given \hat\theta, \mathcal{H}_1)}}_{\text{Likelihood ratio} \; \mathcal{L}_0 / \mathcal{L}_1} \frac{\text{p}(\hat\theta \given y, \mathcal{H}_1)}{g(\hat\theta \given \mathcal{H}_1)}.
\end{aligned}
\]

Jeffreys's approximation is obtained by replacing the posterior density
by a normal distribution. Asymptotic Bayesian theory establishes that
(under suitable regularity conditions) the asymptotic posterior
distribution of \(\theta\) is normal with mean equal to the MLE
\(\hat\theta\) and variance equal to the squared standard error
\(\text{SE}(\hat\theta)^2\) (\citeproc{ref-Held_2020}{Held \& Bové,
2020, p. 204} ff.). If the posterior is approximately
\(\mathcal N(\hat\theta, \mathrm{SE}(\hat\theta)^2)\), then

\[
\begin{aligned}
\text{BF}_{01} & = \frac{\mathcal{L}_0}{\mathcal{L}_1} \; \frac{\left\{ \sqrt{2 \pi} \; \text{SE}(\hat\theta) \right\}^{-1}}{g(\hat\theta \given \mathcal{H}_1)} \\
& \approx \frac{1}{\sqrt{2 \pi} \; \text{SE}(\hat\theta) \; g(\hat\theta \given \mathcal{H}_1)} \; \exp\left( -0.5 \; W \right) \\
& = \frac{1}{\sqrt{2 \pi} \; \sigma \; g(\hat\theta \given \mathcal{H}_1)} \; \sqrt{n_\text{eff}} \; \exp\left( -0.5 \; W \right) \\
& = A \; \sqrt{n_\text{eff}} \; \exp\left( -0.5 \; W \right) \\
& = \text{JAB}_{01},
\end{aligned}
\] because the Wald-statistic \(W\) is asymptotically equivalent to
\(2 \, \log(\mathcal{L}_1 / \mathcal{L}_0)\) under standard regularity
conditions (\citeproc{ref-Buse_1982}{Buse, 1982};
\citeproc{ref-Engle1984}{Engle, 1984, p. 798};
\citeproc{ref-Vaart_1998}{Vaart, 1998, p. 227}).

\subsection{Taylor series expansion}\label{taylor-series-expansion}

While the Dirac limit justifies the approximation at infinity and Chib's
identity provides a justification for finite data, both rely on the
assumption that the likelihood overwhelms the prior. To explore the
conditions under which this is the case, we can use a Taylor series
expansion of the prior (\citeproc{ref-Gelman2013}{Gelman et al., 2013,
pp. 83--84}). This approach explores the mechanics of the marginal
likelihood integral for large but finite samples, i.e., when
\(\text{SE}(\hat\theta)\) is small.

Assuming that the prior distribution is at least twice continuously
differentiable at \(\hat\theta\), the Taylor series around
\(\hat\theta\) expands to

\[
g(\theta \given \mathcal{H}_1) = g(\hat\theta \given \mathcal{H}_1) + g'(\hat\theta \given \mathcal{H}_1) (\theta - \hat\theta) + \frac{1}{2} g''(\hat\theta \given \mathcal{H}_1) (\theta - \hat\theta)^2 + ...,
\] where \(g'()\) and \(g''()\) are the first and second derivative of
the prior distribution. Note that the likelihood can be rewritten as
\(\mathcal{N}(\hat\theta \given \theta, \text{SE}(\hat\theta)^2) = \mathcal{N}(\theta \given \hat\theta, \text{SE}(\hat\theta)^2)\).
Substituting the Taylor expansion into Equation~\ref{eq-ml-approx} and
evaluating the terms using the moments of the normal distribution yields

\[
\begin{aligned}
\text{p}(\hat\theta \mid \mathcal{H}_1) = & \int \mathcal{N}(\theta \mid \hat\theta, \text{SE}(\hat\theta)^2) \left[ g(\hat\theta \given \mathcal{H}_1) + g'(\hat\theta \given \mathcal{H}_1) (\theta - \hat\theta) + \frac{1}{2} g''(\hat\theta \given \mathcal{H}_1) (\theta - \hat\theta)^2 + \cdots \right] d\theta \\
= & \; g(\hat\theta \given \mathcal{H}_1) \int \mathcal{N}(\theta \mid \hat\theta, \text{SE}(\hat\theta)^2) \; d\theta \; + \\
& g'(\hat\theta \given \mathcal{H}_1) \int \mathcal{N}(\theta \mid \hat\theta, \text{SE}(\hat\theta)^2) \; (\theta - \hat\theta) \; d\theta \; + \\
& \frac{1}{2} \; g''(\hat\theta \given \mathcal{H}_1) \int \mathcal{N}(\theta \mid \hat\theta, \text{SE}(\hat\theta)^2) \; (\theta - \hat\theta)^2 \; d\theta + \cdots.
\end{aligned}
\]

The first integral
\(\int \mathcal{N}(\theta \mid \hat\theta, \text{SE}(\hat\theta)^2) \, d\theta = 1\).
The second and third integrals correspond to the first and second
central moments of the normal distribution. Hence
\(\int (\theta - \hat\theta) \; \mathcal{N}(\theta \mid \hat\theta, \text{SE}(\hat\theta)^2) \, d\theta = 0\),
and
\(\int \mathcal{N}(\theta \mid \hat\theta, \text{SE}(\hat\theta)^2) \; (\theta - \hat\theta)^2 \; d\theta = \text{SE}(\hat\theta)^2\)
and therefore

\[
\text{p}(\hat\theta \mid \mathcal{H}_1) = g(\hat\theta \given \mathcal{H}_1) \; + \frac{1}{2} \; g''(\hat\theta \given \mathcal{H}_1) \; \text{SE}(\hat\theta)^2 + \cdots.
\]

This result shows that the approximation is accurate if
\(g(\hat\theta \given \mathcal{H}_1)\) dominates the sum, either because
the standard error \(\text{SE}(\hat\theta)\) is small or because the
prior distribution is approximately constant in the region where the
likelihood is concentrated, i.e.,
\(g''(\hat\theta \given \mathcal{H}_1)\) is small. For a location-scale
prior with scale \(\sigma_\theta\), this typically means that
\(\sigma_\theta\) is large compared with \(\mathrm{SE}(\hat\theta)\), so
the prior is locally flat over the region where the likelihood is
concentrated, \(\text{SE}(\hat\theta) \ll \sigma_\theta\).

\section{Minimum Effective Sample Size to Limit JAB
Bias}\label{sec-min-n-bias}

To solve Equation~\ref{eq-upi-jab-bias} for \(n_\text{eff}\), we define
the bias \(k = \text{BF}_{01} / \text{JAB}_{01}\), the standardized
effect size \(\hat\delta = (\hat\theta - \theta_0) / \sigma\), note that
\(1 + 1/ n_\text{eff} = (n_\text{eff} + 1) / n_\text{eff}\), and square
the bias equation:

\[
k^2 = \frac{n_{\text{eff}} + 1}{n_{\text{eff}}} \exp \left( -\frac{\hat\delta^2}{n_{\text{eff}} + 1} \right).
\] Next, we define

\[
u = \frac{n_{\text{eff}}}{n_{\text{eff}} + 1} \quad \implies \quad 1 - u = \frac{1}{n_{\text{eff}} + 1},
\] substitute \(u\) into the squared equation, isolate terms involving
\(u\), and multiply both sides by \(-\hat\delta^2\),

\[
\begin{aligned}
k^2 & = \frac{1}{u} \exp\left\{ -\hat\delta^2(1 - u) \right\} \\
u k^2 & = \exp(-\hat\delta^2) \exp(\hat\delta^2 u) \\
u \exp(-\hat\delta^2 u) & = \frac{1}{k^2} \exp(-\hat\delta^2) \\
(-\hat\delta^2 u) \exp(-\hat\delta^2 u) & = -\frac{\hat\delta^2}{k^2} \exp(-\hat\delta^2).
\end{aligned}
\] This yields the standard form \(Y e^Y = X\) to apply the Lambert
\(\mathcal{W}\) function by multiplying both sides by \(-\hat\delta^2\),

\[
-\hat\delta^2 u = \mathcal{W}\left\{ -\frac{\hat\delta^2}{k^2} \exp\left( -\hat\delta^2 \right) \right\}.
\]

To isolate \(n_\text{eff}\), we define
\(C = \mathcal{W}\left\{ -\frac{\hat\delta^2}{k^2} \exp\left(-\hat\delta^2\right) \right\}\)
and substitute \(u = \frac{n_{\text{eff}}}{n_{\text{eff}} + 1}\) back
into the equation,

\begin{align*}
-\hat\delta^2 \left( \frac{n_{\text{eff}}}{n_{\text{eff}} + 1} \right) &= C \\
-\hat\delta^2 n_{\text{eff}} &= C n_{\text{eff}} + C \\
n_{\text{eff}}(-\hat\delta^2 - C) &= C \\
n_{\text{eff}} &= \frac{-C}{\hat\delta^2 + C}.
\end{align*}
This yields closed-form solution

\[
n_{\text{eff}} = \frac{-\mathcal{W}\left\{ -\frac{\hat\delta^2}{k^2} \exp\left(-\hat\delta^2\right) \right\}}{\hat\delta^2 + \mathcal{W}\left\{ -\frac{\hat\delta^2}{k^2} \exp\left(-\hat\delta^2\right) \right\}}.
\]

To better understand the interplay of effective sample size,
standardized effect size, and bias, we can derive an approximate
solution using a first-order Taylor expansion at \(1/n_\text{eff} = 0\).
For the natural logarithm of the squared bias equation,

\[
2 \log(k) = \ln\left(1 + \frac{1}{n_{\text{eff}}}\right) - \frac{\hat\delta^2}{n_{\text{eff}} + 1},
\] we apply the two first-order Taylor approximations
\(\log(1 + 1/n_{\text{eff}}) \approx \frac{1}{n_{\text{eff}}}\) and
\(\frac{1}{n_{\text{eff}} + 1} \approx \frac{1}{n_{\text{eff}}}\).
Substituting these into the above equation yields

\[
\begin{aligned}
2 \log(k) & \approx \frac{1}{n_{\text{eff}}} - \frac{\hat\delta^2}{n_{\text{eff}}} \\
n_{\text{eff}} & \approx \frac{1 - \hat\delta^2}{2 \ln(k)}.
\end{aligned}
\]

\end{document}